\documentclass[]{interact}

\usepackage{epstopdf}
\usepackage[caption=false]{subfig}

\usepackage{float}

\usepackage{txfonts}
\usepackage{color}
\usepackage{colortbl}
\usepackage{multirow}
\usepackage[table,xcdraw]{xcolor}

\usepackage{enumitem}
\setlist{leftmargin=1.25cm}

\theoremstyle{plain}
\newtheorem{theorem}{Theorem}[section]

\newtheorem{proposition}[theorem]{Proposition}

\theoremstyle{definition}

\theoremstyle{remark}
\newtheorem{remark}{Remark}

\begin{document}


\title{Quadratic estimation for stochastic systems in the presence of random parameter matrices, time-correlated additive noise and  deception attacks}

\author{
\name{R. Caballero-{\'A}guila\textsuperscript{a}\thanks{This is preprint submitted to Journal of the Franklin Institute on February 15, 2023, available at: https://doi.org/10.1016/j.jfranklin.2023.08.033.},
and
J. Linares-P{\'e}rez\textsuperscript{b}}
\affil{\textsuperscript{a}Departamento de Estad\'istica. Universidad de Ja{\'e}n.  Paraje Las Lagunillas. 23071. Ja{\'e}n. Spain; \\ \textsuperscript{b}Departamento de Estad\'istica. Universidad de Granada. Avenida Fuentenueva. 18071. Granada. Spain}
}

\maketitle

\begin{abstract}
Networked systems usually face different random uncertainties that make the performance of the least-squares (LS) linear filter decline significantly. For this reason, great attention has been paid to the search for other kinds of suboptimal estimators. Among them, the LS quadratic estimation approach has attracted considerable interest in the scientific community for its balance between computational complexity and estimation accuracy. {When it comes to stochastic systems subject to different random uncertainties and deception attacks, the quadratic estimator design has not been deeply studied.} In this paper, using covariance information, the LS quadratic { filtering and fixed-point smoothing problems are}
addressed under the assumption that the measurements are perturbed by a
time-correlated additive noise, as well as affected by random parameter matrices
and exposed to random deception attacks. { The use of random parameter matrices covers a wide range of
common uncertainties and random failures, thus better reflecting the engineering reality.} The signal and observation
vectors are augmented by stacking the original vectors with their second-order
Kronecker powers; then, the linear estimator of the original signal
based on the augmented observations provides the required quadratic estimator.
A simulation example illustrates the superiority of the proposed
quadratic estimators over the conventional linear ones { and the effect of the deception attacks on the estimation performance}.
\end{abstract}


%
%
%




\section{Introduction}
\label{sec:introd}

Sensor networks are becoming increasingly popular in a broad range of application fields, including health care, military, transportation, mining, agriculture, intelligent buildings and smart cities, among others  \cite{Singh_et_al_Springer_2022}. As a result, the fusion estimation problem in networked systems is attracting tremendous research interest.

However, it must be mentioned that communication networks usually suffer resource constraints and, consequently, some network-induced phenomena will inevitably emerge during signal measurement or transmission \cite{Liu_et_al_Springer_2019}. Some of the most common networked-induced phenomena occurring in different application disciplines --presence of multiplicative noise, missing observations, or fading measurements, among others-- can be globally characterized by incorporating stochastic parameters in the measurement equations. {
 Thus, the use of random parameter matrices in the measurement equation allows us to model the randomness in the measurements and to account for different uncertainties in many real-life scenarios --e.g., radar systems, wireless communication, sensor networks or environmental monitoring-- where such uncertainties can occur.} As a result, research on the estimation problem in systems with random parameter matrices has grown in popularity during the past few years. See, for example, \cite{Yang_2016}-\cite{Caballero_IJSS_2023} and the references therein for some sample contributions.

The traditional Kalman-type filtering problem is usually based on the assumption that the measurement additive noise is either white or finite-step correlated. In practice, however,  infinite-step correlated measurement noises can be found in a wide variety of engineering applications, where the sampling frequency is typically high enough to cause measurement noises to be significantly correlated over two or more consecutive sampling periods. Over the last decade, numerous papers have addressed the estimation problem under the assumption that the infinite-step time-correlated channel noise is the output of a linear system model with white noise.  The state augmentation method --which is simple and direct but computationally expensive-- and the measurement differencing method --which avoids increasing dimensions,  but requires two consecutive measurements to compute the difference-- are the most popular methods for dealing with this type of noise correlation (see, e.g., \cite{Li_2017}-\cite{Caballero_INFFUS_2020}). More recently, alternative non-augmentation methods that do not require the availability of consecutive measurements have been proposed to address the state estimation problem in linear systems with time-correlated additive noises and random packet dropouts in \cite{Ma_SIGPRO_2020} and \cite{Caballero_SENSORS_2022}.

When dealing with the estimation problem in networked systems, security is an important topic that should not be overlooked. The possibility of suffering cyber attacks is one of the most common weaknesses (see, e.g., \cite{Mahmoud_2019} and \cite{Sanchez_2019}) and, particularly, the estimation problem in networked systems subject to deception attacks have inspired many significant research studies. Generally speaking, the major goal of deception attackers is altering the data integrity by maliciously falsifying their information in a random way. The centralized security-guaranteed filtering problem is studied for linear time-invariant stochastic systems with multirate-sensor fusion under deception attacks in \cite{Wang_2018_IJFI}. The H$_\infty$-consensus filtering problem for discrete-time systems with multiplicative noises and deception attacks is investigated in \cite{Han_2019} {
and the chance-constrained $H_\infty$ state estimation problem is investigated for a class of time-varying neural networks subject to measurements degradation and randomly occurring deception attacks in \cite{Qu_2022}.}
The distributed estimation problem in sensor networks with a specific topology structure has been studied in \cite{Yang-et-al-2019-Automatica} --under false data injection attacks-- and in \cite{Caballero_SENSORS_2020}-\cite{Ma-et-al-IEEE_TSIPN_2021} --under deception attacks.


It is unquestionable that the LS estimation problem of random signals from noisy measurements has played a key role over the past decades. The well-known Kalman filter provides the LS signal estimator for linear systems subject to Gaussian, mutually independent initial signal and noise processes. However, in the presence of non-Gaussian disturbances, only the LS linear estimator is provided and the optimal estimator is, in general, computationally expensive. Besides, due to the network-induced random uncertainties described in the previous paragraphs, networked systems are generally non-Gaussian and the practical computation of the LS estimator usually involves a significant complexity. For this reason, a great deal of attention has been devoted to the design of simpler suboptimal estimators with satisfactory accuracy, being the design of LS linear estimation algorithms the most popular approach. A more effective scheme to address the estimation problem in non-Gaussian systems is the LS quadratic approach, due to its usual outperformance over the LS linear one and its adequate balance between computational burden and estimation accuracy.  In linear discrete-time non-Gaussian systems, the input noise quadratic estimation problem is addressed in \cite{Zhao_Zhang-Neurocomputing2016}
 and a recursive quadratic estimation algorithm for the system state is proposed in \cite{Caballero_et_al_AMC2016} under the presence of random parameter matrices.
 A feedback quadratic filtering algorithm, that reduces the estimation error with respect to the plain quadratic filter, is proposed in \cite{Cacace_et_al-Automatica2017}. The quadratic estimation problem has also been addressed in discrete-time systems with measurement delays and packet dropouts \cite{Li_et_al-Complexity2020} and under the presence of multiplicative noises and quantization effects \cite{Liu_et_al-Automatica2020}. Recently, recursive quadratic estimation algorithms are proposed in \cite{Wang_et_al-IEEETSP2022} --for linear systems over time-correlated fading channels-- and \cite{Wang_et_al-Automatica2023} --for nonlinear systems with energy-harvesting sensors. However, to the best of the authors' knowledge, there have been scarce studies on the quadratic estimation problem in linear systems with random parameter matrices and time-correlated additive noise, let alone the scenario where random attacks are also involved.

{ Inspired by the discussion made so far, our aim is to address the LS quadratic filtering and fixed-point smoothing estimation problems for a class of stochastic systems in the presence of random parameter matrices, time-correlated additive noise and random deception attacks. The following are the key difficulties we are dealing with: {\em (1)} Analysis of the statistical properties of the augmented noises produced by the original time-correlated additive noises and their Kronecker products. {\em (2)} Development of an effective quadratic estimation method in the presence of random parameter matrices, time-correlated additive noise and random deception attacks. {\em (3)} Evaluation of the impact of deception attacks on estimation performance.}

{ The main contributions of this paper are summarized as follows:
{\em (a)} The class of systems investigated in this paper is quite comprehensive, as the use of random parameter matrices embraces a wide range of
common uncertainties and random failures, thus better reflecting the technical reality. {\em (b)} The original observation vectors are augmented with their second-order Kronecker powers, so the quadratic estimation problem is reformulated as a linear estimation problem from the augmented observations and recursive formulas for the estimation error covariances are also proposed. {\em (c)} A covariance-based estimation approach is used, so the evolution model of the signal to be estimated does not need to be known.  {\em (d)}  The direct estimation of the time-correlated additive
noise avoids the use of the differencing method.  {\em (e)} The proposed LS quadratic filtering and fixed-point smoothing estimators outperform the conventional linear ones.
}

\vskip .25cm

The paper is organized as follows. The characteristics of the observation model under consideration are described in Section 2. The LS quadratic estimation problem is formulated in Section 3, where the augmented vectors are defined. The study of the dynamics of the augmented vectors (subsection 3.1) and their second-order statistical properties (subsection 3.2) will be the key to obtain a new observation model, from which the recursive quadratic estimation algorithms are derived in Section 4. A simulation study, in Section 5, shows the effectiveness of the proposed filtering and fixed-point smoothing estimators, as well as their superiority over the conventional linear ones. Finally, some conclusions are given in Section 6, which is followed by three appendices that provide the mathematical proofs of the main theoretical results.

\vskip.25cm
\emph{Notation.} As far as possible, standard mathematical notation will be used throughout the paper. If not explicitly stated, the dimension of all vectors and matrices is assumed to be compatible with algebraic operations.

\begin{table}[H]
{\scriptsize
\begin{tabular}{ll}
\hline
$\mathbb{R}^{n}$ & Set of $n$-dimensional real vectors\\
\rowcolor[HTML]{EFEFEF}
$\delta_{k,h}$ &  Kronecker delta function\\
$M^T$ and $M^{-1}$ &   Transpose and inverse of   matrix $M$\\
\rowcolor[HTML]{EFEFEF}
$M^{(a)T}$ and $M^{(a)-1}$ & Shorthand for $(M^{(a)})^T$ and $(M^{(a)})^{-1}$ \\
$(M_1 \ | \  \ldots \ | \   M_k)$ &   Partitioned matrix whose blocks are the submatrices $M_1, \ldots, M_k$\\
\rowcolor[HTML]{EFEFEF}
$Diag(N_1,\ldots, N_m)$& Block diagonal matrix with main-diagonal blocks $N_1, \ldots, N_m$\\
$0$ & Zero scalar or matrix of compatible dimension\\
\rowcolor[HTML]{EFEFEF}
$\otimes$ & Kronecker product of matrices \\
$K_{n^{2}}$ & $n^2\times n^2$  matrix such that $K_{n^{2}}(z\otimes  v)=v\otimes z + z \otimes v, \forall  z,v \in \mathbb{R}^n$\\
\rowcolor[HTML]{EFEFEF}
$M^{[2]}= M \otimes M$  & Second-order Kronecker power of vector or matrix $M$\\
$vec(\star)$ & $vec$ operator \\
\rowcolor[HTML]{EFEFEF}
$E[a]=\overline{a}$ &  Mathematical expectation of a random vector or matrix $a$\\
$G_{k}=G_{k,k}$ & Function $G_{k,h}$, depending on time instants $k$ and $h$, when $h=k$\\
\rowcolor[HTML]{EFEFEF}
$\Sigma_{k,s}^{{ab}}$&  Covariance  of  random vectors $a_k$ and $b_s$  ($\Sigma_{k,s}^{a}=\Sigma_{k,s}^{aa}$)\\
\rowcolor[HTML]{EFEFEF}
&  $\Sigma_{k,s}^{ab}=Cov[a_k,b_s]=E\big[\big(a_k-\overline{a}_k\big)\big(b_s-\overline{b}_s\big)^T\big], \ \ Cov[a_k]=\Sigma_{k}^{a}$ \\
$\widehat{a}_{k/s}$& Optimal quadratic estimator of the vector $a_k$ based on $\big\{ y_1,\ldots,y_s\big\}$\\
\cline{1-2}
\end{tabular}
}
\end{table}

\section{Observation model}
\label{sec:observ_model}

Consider a random signal, $x_k \in \Bbb R^{n_x}$, to be estimated and assume that the actual measurements, $z_k \in \Bbb R^{n_z}$,  are described by
\begin{equation}\label{actual_measurements}
  z_k = H_k x_k + v_k, \ \ k\geq 1,
\end{equation}
where $H_k$ are random parameter matrices and $v_k$ is a time-correlated additive noise, satisfying
\begin{equation}\label{time-correlated_noise}
  v_k=D_{k-1}v_{k-1}+u_{k-1}, \ \ k\geq 1,
\end{equation}
in which $D_k$ are known, non-singular, time-varying matrices and $u_k$ is a white noise.

{ \begin{remark} Usually, in physical electronic systems, white noise becomes time-correlated when it passes through bandlimited channels. When the noise has a correlation time that is significantly shorter than the relevant time intervals of interest, it is generally regarded as white noise, and its colored nature is typically disregarded. However, if the correlation time is not negligible, the system may encounter significant interference from the colored noise. Model (2) is appropriate to simulate, for example, the signal strength colored noise in Global Positioning System (GPS) receivers or the GPS positioning noise due to carrier phase multipath errors (see \cite{Shmaliy2020} and references therein).
\end{remark}
}

{ Let us assume that deception attacks are launched by an adversary, who injects a false signal  modeled by}
\begin{equation}\label{deception_attack}
\breve{z}_k=-z_k+w_k, \ \ k\geq 1.
\end{equation}

At each sampling time $k$, the attack can randomly succeed or fail and this fact is described by a Bernoulli random variable, $\lambda_k$, whose values --one or zero-- represent a successful or failed attack, respectively. Therefore, the available observations, $y_k$, that will be used to estimate the signal,  are given by $y_k=z_k+ \lambda_k \breve{z}_k$ or, equivalently, using (\ref{deception_attack}),
\begin{equation}\label{available_observ}
y_k=(1-\lambda_k)z_k+\lambda_kw_k, \ \ k\geq 1.
\end{equation}

{ \begin{remark}
According to (3), the false data injected  by the attackers, $\breve{z}_k = -z_k+w_k$, are assumed to be divided mathematically into two components: a neutralizing one, $-z_k$, that will cancel the original measurement and a noise component, $w_k$, that represents the blurred deceptive information added by the attacker. Hence, at each instant of time, the compromised measurement (4) can be the actual measurement (if the attack fails) or only noise (if the attack succeeds).
Another interesting kind of deception signal could be, for example, $\breve{z}_k = w_k$, which involves adding random noise to the received measurement. This kind of deception attacks aims to degrade the measurement quality by introducing unwanted noise. Unlike (4), under this model, the resulting compromised measurements are given by $y_k=z_k+\lambda_k w_k$, thus always containing the actual measurement (with or without noise, depending on whether the attack is successful or not, respectively).
A complete survey of different kinds of attacks can be seen, for example, in \cite{Mahmoud_2019}.
\end{remark}}

{ \begin{remark}
The mathematical model of the compromised measurements (4) looks similar to the packet loss model. The main difference between both models lies in the noise component, $w_k$, that represents the
blurred deceptive information added by the attacker.
In the random packet loss scenario, at each time instant, the processing center can receive the actual measurement (if there is no loss) or nothing (if the actual packet is lost), in which case it is usually compensated with either the most recently received packet or the prediction estimate of the lost measurement, thus always providing valuable information for the signal estimation. However, the possibility of random deception attacks means that, at each time instant, the processing center can receive the actual measurement (if the attack fails) or only noise (if the attack succeeds). Hence, in addition to neutralizing the original measurement, what could be mathematically similar to a packet dropout, the deception attack degrades the quality of the measurements by introducing unwanted noise, thus adding further difficulties in obtaining accurate estimation algorithms.
\end{remark}}

Taking into account the difficulty in obtaining the least-squares (LS) optimal estimator in the presence of random uncertainties in the measurements, most studies are focused on the LS linear estimation problem.
As it is well-known, when the random processes involved in the observation model have finite second-order moments, the LS linear estimator of $x_k$ given the observations $y_1, \ldots, y_L$ is the orthogonal projection of $x_k$ on the space of $n_x$-dimensional random vectors produced as linear transformations of such observations. Nonetheless, some efforts have also been directed towards the search for estimation algorithms that keep the advantages of the linear ones --recursivity and computational simplicity-- and provide more precise estimates.  {Specifically, our aim is to design recursive algorithms to obtain LS quadratic estimators of the signal $x_k$ given the observations $y_1,\ldots,y_L$;} such estimators are given by the orthogonal projection of the vector $x_k$ onto the linear space generated by the observations $y_1, \ldots, y_L$  and their second-order Kronecker powers $y_1^{[2]}, \ldots, y_L^{[2]}$.
Therefore, to address the LS quadratic estimation problem, it is necessary that the second-order moments of the Kronecker powers $y_1^{[2]}, \ldots, y_L^{[2]}$ exist, for which all the random processes involved must be fourth-order processes and we will assume them to have known finite second, third and fourth moments.

\vskip .15cm
{ To simplify the statement of the subsequent assumptions and properties of the processes involved, we introduce the following notations for the covariance and cross-covariance functions of a stochastic process $\big\{a_k\big\}_{ k\geq1}$ and its second-order Kronecker powers:
\vskip-.75cm$$\begin{array}{l}\!\!\Sigma^{a}_{k,s}\!\!= Cov[a_k, a_s], \ \Sigma^{a^{(2)}}_{k,s}\!\!= Cov[a^{[2]}_k, a^{[2]}_s], \ \Sigma^{a^{(12)}}_{k,s}\!\!\!= Cov[a_k, a^{[2]}_s], \ \Sigma^{a^{(2\hskip .04cm 1)}}_{k,s}\!\!\!=\Sigma^{a^{(12)T}}_{s,k}\hskip-.25cm.\\
\!\!\Sigma^{a}_{k}= Cov[a_k], \ \Sigma^{a^{(2)}}_{k}\!\!= Cov[a^{[2]}_k], \ \Sigma^{a^{(12)}}_{k}\!\!\!= Cov[a_k, a^{[2]}_k], \ \Sigma^{a^{(2\hskip .04cm 1)}}_{k}\!\!\!=\Sigma^{a^{(12)T}}_{k}\hskip-.5cm.\end{array}$$

The following assumptions are made:
}
\begin{itemize} [leftmargin=\dimexpr 26pt-.00in]

 \item[\em (H1)]
  The  signal process $\big\{x_k\big\}_{ k\geq1}$  has    zero mean and
its covariance function, $  \Sigma^x_{k,s}$, as well as the
covariance function of its second-order powers,  $\Sigma^{x^{(2)}}_{k,s}$, can be factorized as follows:
    $$  \Sigma^x_{k,s}=  A_kB^T_s,\ \ \ \Sigma^{x^{(2)}}_{k,s}=\breve{A}_k \breve{B}^{T}_s, \  s\leq k,$$
 where the $n_x\times M_1$ matrices   $A_k$, $B_s$ and the $n_x^2\times M_2$ matrices $\breve{A}_k$, $\breve{B}_s$ are  known  for all $k,s\geq 1$.  Also, the cross-covariance function  of the signal and its second-order powers,
$\Sigma^{x^{(12)}}_{k,s}$, can be expressed as:
{\[ \Sigma^{x^{(12)}}_{k,s}=
\left\{
\begin{array}[c]{l}
A_{1,k} {B}^T_{1,s},\ \ \  s\leq k,  \\
{B}_{2,k} A^T_{2,s},\ \ \ \ k\leq s,
\end{array}
\right.
\]}
where, for all $k,s\geq 1$,  { $A_{1,k}$, $B_{1,s}$,  $B_{2,k}$ and $A_{2,k}$} are   $n_x\times P_1$, $n_x^2\times P_1$,
$n_x\times P_2$ and $n_x^{2}\times P_2$ known matrices, respectively.

 \item[\em (H2)] $\big\{H_k\big\}_{ k\geq1}$ is a sequence of independent random parameter matrices with known
mean matrices $\overline{H}_{k}$. The covariances and cross-covariances between the entries  of the
matrices  $H_k$ and $H_k^{[2]}$, are also assumed to be
known.

 \item[\em (H3)]  $v_{0}$ is a zero-mean random vector whose moments up to
the fourth-order one  are known.

 \item[\em (H4)] The noise processes $\{u_{k}\}_{k\geq 0}$ and $\{w_{k}\}_{k\geq 1}$ are zero-mean white sequences with known moments, up to the fourth-order  ones.

 \item[\em (H5)]
 $ \big\{\lambda_k\big\}_{ k\geq 1}$,   is a sequence of  independent  Bernoulli random variables with  known probabilities $P\big(\lambda_k=1\big)=\overline{\lambda}_k, \  k\geq 1.$

\item[\em (H6)] The signal process $\{x_k\}_{ k\geq1}$, the vector $v_0$  and the processes  $\{H_k\}_{ k\geq1}$, $\{u_k\}_{ k\geq0}$,
  $\{w_k\}_{ k\geq 1}$  and $\{\lambda_k\}_{ k\geq1}$
      are mutually independent.

\end{itemize}

\vskip .25cm

\begin{remark}
The derivation of the proposed quadratic estimation  algorithms will not  require the evolution model of the signal; instead, we will use a  covariance-based estimation approach. In this approach, although  the signal evolution model is not necessary, a zero-mean signal is required and the covariance and cross-covariance functions of the signal and its second-order powers are to be expressed in a separable form.
It should be noted that these assumptions, imposed in (H1), are met under the most commonly used signal evolution models.
 For instance,  let us consider a zero-mean non-stationary signal obeying a linear evolution model $x_k=\phi_{k-1}x_{k-1}+\varepsilon_{k-1}, \ k\geq 1$, with non-singular transition matrices, $\phi_{k}$, and the following assumptions:
 \begin{itemize}[leftmargin=\dimexpr 26pt-.1in]
  \item[$\circ$] $x_{0}$ is a zero-mean random vector whose moments, up to
the fourth-order one, are  known.

  \item[$\circ$]  $\{\varepsilon_{k}\}_{k\geq 0}$ is a zero-mean white process with known moments, up to the fourth-order ones.

\end{itemize}
The  covariance and cross-covariance functions of the signal and their second-order powers  are given by
\vskip-.75cm $$\Sigma^x_{k,s}= \phi_{k,s}\Sigma_s^x,  \ \  \Sigma^{x^{(2)}}_{k,s}= \phi^{[2]}_{k,s}\Sigma^{x^{(2)}}_{s}, \ s\leq k;
 \ \ \ \Sigma^{x^{(12)}}_{k,s}=
\left\{
\begin{array}[c]{l}
\phi_{k,s}\Sigma^{x^{(12)}}_{s},\ \  s\leq k,  \\
\Sigma^{x^{(1 2)}}_{k}\phi^{[2]T}_{s,k},\  \ k\leq s,
\end{array}
\right.$$
\vskip-.25cm \noindent where $\phi_{k,s}=\phi_{k-1}\cdots \phi_s$, $\phi^{[2]}_{k,s}=\phi^{[2]}_{k-1}\cdots \phi^{[2]}_s$ and the functions $\Sigma^x_s$, $\Sigma^{x^{(12)}}_{s}$ and $\Sigma^{x^{(2)}}_{s}$ are recursively obtained by
$$\begin{array}{l}\Sigma_s^x = \phi_{s-1} \Sigma_{s-1}^x \phi_{s-1}^T + \Sigma^{\varepsilon}_{s-1},\  \ s\geq 1,
 \\
 \Sigma^{x^{(12)}}_{s} = \phi_{s-1} \Sigma_{s-1}^{x^{(12)}} \phi_{s-1}^{[2]T} + \Sigma^{\varepsilon^{(12)}}_{s-1},\  \ s\geq 1,
 \\
\Sigma_s^{x^{(2)}} = \phi^{[2]}_{s-1} \Sigma_{s-1}^{x^{(2)}} \phi_{s-1}^{[2]T} + K_{n^2_x}\big(\phi_{s-1}\Sigma^x_{s-1}\phi^T_{s-1}
 \otimes \Sigma^\varepsilon_{s-1}\big)K_{n^2_x}+\Sigma^{\varepsilon^{(2)}}_{s-1},\  \ s\geq 1.
 \end{array}$$
 Clearly,  according to assumption  (H1), these covariance functions can be expressed in a separable form  taking, for example, the following functions:
$$\begin{array}{c}A_k= \phi_{k,0}, \ B^T_s= \phi^{-1}_{s,0}\Sigma^x_s; \ \ \ \breve{A}_k= \phi^{[2]}_{k,0}, \ \breve{B}^T_s= (\phi^{[2]}_{s,0})^{-1}\Sigma^{x^{(2)}}_k;\\\\
A_{1,k}= \phi_{k,0}, \  B^T_{1,s}=\phi^{-1}_{s,0}\Sigma^{x^{(12)}}_s; \ \ \ B_{2,k}=\Sigma^{x^{(12)}}_k(\phi^{[2]T}_{k,0})^{-1}, \ A^T_{2,s}= \phi^{[2]T}_{s,0}.\end{array}$$
An analogous reasoning can be carried out for  a stationary signal, as it will be
shown in Section \ref{sec:simul_study}; hence, the separability assumptions on the signal  required in   (H1) covers different types of stationary and non-stationary signals and the estimation based on such hypotheses, instead of the state-space model, provides a unifying context to obtain general algorithms which are applicable to a large number of practical situations.
\end{remark}

\begin{remark}

As already indicated in the Introduction, in order to deal with the time-correlated noise, we will not use  the measurement differencing method, but we will instead address the
direct estimation of the noise. Since the derivation of the estimation algorithms will be carried out  under a covariance-based approach, it is necessary to express the noise covariance function in a separable form.
\vskip .1cm
From (\ref{time-correlated_noise}) and assumptions (H3) and (H4), we have that the covariance function $\Sigma_{k,s}^v$ of the time-correlated noise, $v_k$, is factorized in a separable form $\Sigma_{k,s}^v = \mathbf{D}_k \mathbf{F}_s^T$,   in which
      $\mathbf{D}_k = D_{k,0}, $
      $\mathbf{F}_s^T = D_{s,0}^{-1} \Sigma_s^v$,  where $D_{k,0}=D_{k-1} \cdots D_0$ and $\Sigma_s^v$ is recursively computed by
$$\Sigma_s^v = D_{s-1} \Sigma_{s-1}^v D_{s-1}^T + \Sigma^{u}_{s-1},\  s\geq 1.$$
\end{remark}

\section{Least-squares quadratic estimation}
\label{sec:quadratic_estimation}

Given the observation model (\ref{actual_measurements})-(\ref{available_observ}), under hypotheses \emph{(H1)-(H6)}, our aim is finding the LS quadratic estimator, $\widehat{x}_{k/L}$, of the signal, $x_k$, with knowledge of the observation history up to the $L$th sampling time, $y_1,\ldots,y_L$. More specifically, our goal is to construct recursive algorithms for the filter $\widehat{x}_{k/k}$ and the smoother  $\widehat{x}_{k/L}$ , at the arbitrary fixed point $k$, for any $L>k$.

 {
 The filtering algorithm provides estimators of the current signal, $x_k$, based on the measurements up to the present time, $y_1,\ldots,y_k$.  At each time step $k$, the filtering estimator, $\widehat{x}_{k/k}$, is updated  based on the new measurement, $y_k$, and the previous estimator, $\widehat{x}_{k-1/k-1}$.

The fixed-point smoothing algorithm allows us to obtain the estimator of the signal at a fixed time,
given an increasing number of 
posterior available measurements; that is, the fixed-point smoother is used to estimate the signal, $x_k$, at the fixed point $k$, not only based on measurements up to that time, but also using measurements taken beyond it, $k+1, k+2, \cdots.$}

As already indicated, the quadratic estimator is the orthogonal projection of the vector $x_k$ onto the linear space generated by the observations  $y_1, \ldots, y_L$  and the Kronecker powers $y_1^{[2]}, \ldots, y_L^{[2]}$.
The model hypotheses ensure the existence of the second-order moments of these Kronecker powers and, consequently, the existence
of the quadratic estimators $\widehat{x}_{k/L}$ is guaranteed. By combining the original vectors and their second-order powers, the following augmented vectors are defined to derive such estimators:
$$\mathcal{X}_{k}=\left(\!{\footnotesize
    \begin{array}[c]{c}
        x_k  \\
        x^{[2]}_k
    \end{array}
                }\!\right)\!   ,\     \mathcal{Z}_{k}=\left(\!{\footnotesize
    \begin{array}[c]{c}
        z_k  \\
        z^{[2]}_k
    \end{array}
                }\!\right)\!   ,\     \mathcal{V}_{k}=\left(\!{\footnotesize
    \begin{array}[c]{c}
        v_k  \\
        v^{[2]}_k
    \end{array}
                }\!\right)\!   ,\     \mathcal{Y}_{k}=\left(\!{\footnotesize
    \begin{array}[c]{c}
        y_k  \\
        y^{[2]}_k
    \end{array}
                }\!\right)\!   ,\     \mathcal{W}_{k}=\left(\!{\footnotesize
    \begin{array}[c]{c}
        w_k  \\
        w^{[2]}_k
    \end{array}
    } \! \right)\!.$$

Noting that the space of $n_x$-dimensional linear transformations of $y_1,\ldots,y_L$ and $y^{[2]}_1,\ldots,y^{[2]}_L$  is identical to that of linear transformations of $\mathcal{Y}_1, \ldots, \mathcal{Y}_L$, it is obvious that the
LS quadratic estimator of $x_k$ based on $y_1,\ldots,y_L$
 is the
 LS linear estimator of $x_k$ based on $\mathcal{Y}_1, \ldots, \mathcal{Y}_L$. To obtain this linear estimator, firstly, an observation model for the vectors $\mathcal{Y}_L$ will be created by studying the dynamics of the augmented vectors and, secondly, the second-order statistical characteristics of the processes included in this new model are examined.

\subsection{Dynamics of the augmented vectors}
    \label{subsec:dynamics}
\vskip .15cm

By using  (\ref{actual_measurements}) and the Kronecker product properties, the following expression  for $z^{[2]}_k$
is obtained:
$$
z_{k}^{[2]}=H_{k}^{[2]}x_{k}^{[2]}+  K_{n_z^2}\big(H_{k}x_{k}\otimes v_{k}\big)
+v_{k}^{[2]},\quad k\geq 1.
$$
Hence,  the augmented measurement vectors $\mathcal{Z}_k$ clearly satisfy
\begin{equation}\label{augmented actual observ}
\mathcal{Z}_{k}=\mathcal{H}_{k}\mathcal{X}_{k}+\mathcal{V}_{k}+\mathcal{V}^*_{k},\quad k\geq 1,
\end{equation}
in which
$$\mathcal{H}_k=\begin{pmatrix}
H_{k} & 0\\
0     & H_{k}^{[2]}
\end{pmatrix},\ \ \
\mathcal{V}^*_k=\left(\!\!
\begin{array}
[c]{c}
0  \\ K_{n_z^2}\big(H_{k}x_{k}\otimes v_{k}\big)
\end{array}
\!\right).$$

In order to obtain the evolution model for the augmented noise  $\mathcal{V}_k$ in this equation, let us observe that, from (\ref{time-correlated_noise}), it is clear that
$$
v_{k}^{[2]}=D_{k-1}^{[2]}v_{k-1}^{[2]}+  K_{n_z^2}\big(D_{k-1}v_{k-1}\otimes u_{k-1}\big)
+u_{k-1}^{[2]},\quad k\geq 1.
$$
Hence, the augmented noise $\mathcal{V}_k$ is time-correlated and the following equation holds:
\begin{equation}\label{time-correlated augmented noise}
  \mathcal{V}_{k}=\mathcal{D}_{k-1}\mathcal{V}_{k-1}+\mathcal{U}_{k-1},\quad k\geq 1.
\end{equation}
in which
$$\mathcal{D}_k=\begin{pmatrix}
D_{k} & 0\\
0     & D_{k}^{[2]}
\end{pmatrix},
\quad\quad \mathcal{U}_k=\left(\!\!
\begin{array}
[c]{c}
u_k  \\ K_{n_z^2}\big(D_kv_k
 \otimes u_k\big)+
u^{[2]}_k
\end{array}
\!\right).$$

\vskip.25cm

Finally, using (\ref{available_observ}) and taking into account that $\lambda_k(1-\lambda_k)=0$, the augmented observation vectors  $\mathcal{Y}_k$ can be expressed as
\begin{equation}\label{augmented observation}
  \mathcal{Y}_{k}=(1-\lambda_k)\mathcal{Z}_{k}+\lambda_k\mathcal{W}_{k},\quad k\geq 1.
\end{equation}

Note that the augmented vectors in this new model  have non-zero mean. For simplicity in both the study of the statistical properties of the augmented vectors and the derivation of estimation algorithms, the quadratic estimation problem will be addressed by defining the centered vectors:
$$\begin{array}{lll}
\mathbf{x}_k=\mathcal{X}_{k}-\overline{\mathcal{X}}_{k}, \ \      &
\mathbf{z}_k=\mathcal{Z}_{k}-\overline{\mathcal{Z}}_{k}, \ \      &
\mathbf{v}_k=\mathcal{V}_{k}-\overline{\mathcal{V}}_{k}, \ \      \\
\mathbf{y}_k=\mathcal{Y}_{k}-\overline{\mathcal{Y}}_{k}, \ \      &
\mathbf{w}_k=\mathcal{W}_{k}-\overline{\mathcal{W}}_{k}, \ \      &
\mathbf{u}_k=\mathcal{U}_{k}-\overline{\mathcal{U}}_{k}.
\end{array}
$$

Clearly, the LS linear estimator of $x_k$ based on $\mathcal{Y}_{k}, \ldots, \mathcal{Y}_{L}$ is equal to the one based on  $\mathbf{y}_1, \ldots, \mathbf{y}_L$. Theorem \ref{Th dynamics augmented vectors} summarizes the dynamics of the centered augmented vectors (hereafter simply referred to as augmented vectors).

\begin{theorem} \label{Th dynamics augmented vectors}
The dynamics of the augmented vectors are specified as follows:
\begin{itemize}[leftmargin=\dimexpr 26pt-.075in]
  \item[(a)] The augmented measurements $\{\mathbf{z}_k\}_{k\geq 1}$ obey the following equation
    \begin{equation}\label{augmented actual measurement eq}
    {\mathbf{z}}_{k}=\mathcal{H}_{k}{\mathbf{x}}_{k}+{{\mathbf{v}}^{\ast}_{k}+{\mathbf{v}}^{\ast\ast}_k},\quad k\geq 1,
  \end{equation}
where
  $$ { {\mathbf{v}}^{\ast}_k}=  \mathbf{v}_k + \left(\!\!
\begin{array}
[c]{c}
0  \\ K_{n_z^2}\big(\overline{H}_kx_k
 \otimes v_k\big)
\end{array}
\!\right),$$

$$ { {\mathbf{v}}^{\ast\ast}_k}=
\left(\!\!
\begin{array}
[c]{c}
0  \\ \widetilde{H}^{[2]}_k vec(A_kB^T_k) + K_{n_z^2}\big(\widetilde{H}_kx_k
 \otimes v_k\big)
\end{array}
\!\right),$$

\vskip .25cm
\noindent in which
$$\widetilde{H}_k=H_k-\overline{H}_k, \quad \widetilde{H}^{[2]}_k={H}^{[2]}_k-E\big[H^{[2]}_k\big]$$

\vskip .25cm
\noindent and $\{\mathbf{v}_k\}_{k\geq 0}$ is a time-correlated noise, satisfying
\begin{equation}\label{augmented time-correlated noise eq}
  \mathbf{v}_k=\mathcal{D}_{k-1}\mathbf{v}_{k-1}+\mathbf{u}_{k-1},\quad k\geq 1.
\end{equation}

\item[(b)] The following equation holds for the augmented observation process $\{\mathbf{y}_k\}_{k\geq 1}$:
\begin{equation}\label{augmented observation eq}
    \mathbf{y}_{k}=(1-\lambda_k)\mathbf{z}_{k}+\lambda_k \mathbf{w}_{k}-(\lambda_k-\overline{\lambda}_k)\mathbf{g}_k,\quad k\geq 1,
\end{equation}
where $\mathbf{g}_k=\displaystyle\binom{0}{\overline{H}_{k}^{[2]}vec(A_kB^T_k)+vec(\mathbf{D}_k\mathbf{F}^T_k)-vec(\Sigma^w_k)}.$
\end{itemize}

\end{theorem}

\vskip .15cm
\begin{proof}
See Appendix \ref{appendix Proof of Th1}.
\end{proof}

\subsection{Second-order statistical properties of the augmented  processes}
    \label{subsec:properties}

\vskip .15cm
The second-order statistical properties of the augmented processes involved in equations (\ref{augmented actual measurement eq})-(\ref{augmented observation eq}) are set out in the following propositions.


\begin{proposition}\label{proposition 1}

Under hypotheses (H1)-(H6), the processes involved in (\ref{augmented actual measurement eq}) satisfy the following properties:

  \begin{itemize}
   \item[(a)] The augmented signal process $\{ \mathbf{x}_k\}_{k\geq 1}$ has zero mean and its covariance function admits the following factorization:
\begin{equation}\label{Covx}\Sigma^\mathbf{x}_{k,s} = \mathbb{A}_k\mathbb{B}^{T}_s,\ \ \  s\leq k,\end{equation}
\vskip-.5cm
where
\[
\mathbb{A}_k= \left(
\begin{array}
[c]{cccc}
A_k     & A_{1,k}     & 0             & 0\\
0       & 0             & A_{2,k}    & \breve{A}_k
\end{array}
\right),
\
\
\mathbb{B}_k= \left(
\begin{array}
[c]{cccc}
B_k     & 0         &B_{2,k}    & 0    \\
0       & B_{1,k} & 0             & \breve{B}_k
\end{array}
\right).
\]

Also, the expectations $E[x_k\mathbf{x}^{T}_s]$ are factorized as follows:
\begin{equation}\label{Ex}E[x_k\mathbf{x}^{T}_s]=\left\{
\begin{array}
[c]{l}
 \mathbb{\breve{A}}_k\mathbb{B}^{T}_s,\ \ \  s\leq k,  \\
\mathbb{\breve{B}}_k \mathbb{A}^T_s,\ \ \ \ k\leq s,
\end{array}
\right.
\end{equation}
\vskip -.5cm
where
\[
\mathbb{\breve{A}}_k= \left(
A_k  \ | \  A_{1,k}  \ | \  0  \ | \  0
\right),
\
\
\mathbb{\breve{B}}_k= \left(
B_k  \ | \  0  \ | \  B_{2,k}  \ | \   0
\right).
\]

  \item[(b)] The noise process $\{ \mathbf{v}^{\ast}_k\}_{k\geq 1}$  is a sequence of zero-mean random vectors and its covariance function  $\Sigma_{k,s}^{\mathbf{v}^{\ast}}$ can be factorized as follows:
  \begin{equation}\label{Cov-v-ast}\Sigma_{k,s}^{\mathbf{v}^{\ast}}  =  \Delta_k  \Upsilon_s^T, \ s \leq k,\end{equation}
with
$\Delta_k = \left(  \mathbb{C}_k \ | \  \mathbb{D}_k\right),$
$\Upsilon_k = \left(  \mathbb{E}_k \ | \  \mathbb{F}_k\right),    \ \ k\geq  1, $
in which
$$
\begin{array}{cc}
  \mathbb{C}_k= \left(\begin{array}
[c]{c}
0  \\K_{n_z^2}\left(\overline{H}_kA_k \otimes \mathbf{D}_k\right)
\end{array}
\right),
&
\mathbb{D}_k = \mathcal{D}_{k,0}, \\\\
\mathbb{E}_k= \left(\begin{array}
[c]{c}
0  \\K_{n_z^2}\left(\overline{H}_kB_k \otimes \mathbf{F}_k\right)
\end{array}
\right),
&
\mathbb{F}_k =  \Sigma_k^{\mathbf{v}T} \mathcal{D}_{k,0}^{-1T},
\end{array}
$$
where \ $\mathcal{D}_{k,0}=\mathcal{D}_{k-1} \cdots \mathcal{D}_0$ and $\Sigma_s^{\mathbf{v}}$  is recursively computed as follows:
 $$\Sigma_s^{\mathbf{v}} = \mathcal{D}_{s-1} \Sigma_{s-1}^{\mathbf{v}} \mathcal{D}_{s-1}^T + \Sigma^{\mathbf{u}}_{s-1}, \quad  s\geq 1; \quad\quad
      \Sigma_{0}^{\mathbf{v}}=\begin{pmatrix} \Sigma^{v}_{0} & \Sigma^{v^{(12)}}_{0}\\ \Sigma^{v^{(2\hskip .04cm 1)}}_{0}     & \Sigma^{v^{(2)}}_{0} \end{pmatrix}.$$

\item[(c)]  The noise process $\{ {\mathbf{v}}^{\ast\ast}_k\}_{k\geq 1}$ is a sequence of zero-mean mutually uncorrelated random vectors with covariance
 $$\Sigma^{{\mathbf{v}}^{\ast\ast}}_{k}=\begin{pmatrix}
\ 0 & 0\\\\
\ 0 &  {\big(\Sigma^{{\mathbf{v}}^{\ast\ast}}_{k}\big)_{_{22}}}
\end{pmatrix},$$
in which
\begin{equation}\label{22-block}
\begin{array}{ll}
  {\big(\Sigma^{{\mathbf{v}}^{\ast\ast}}_{k}\big)_{_{22}}}= &  E\big[\widetilde{H}^{[2]}_k vec(A_kB^T_k) vec^T(A_kB^T_k) \widetilde{H}^{[2]T}_k\big]  \\
                                     &  + K_{n_z^2}  \Big(E\big[\widetilde{H}_kA_kB^T_k \widetilde{H}_k\big] \otimes \mathbf{D}_k \mathbf{F}_k^T\Big)  K_{n_z^2}.
\end{array}
\end{equation}

\end{itemize}
\end{proposition}
{
\begin{proof}
See Appendix \ref{appendix Proof of Prop1}.
\end{proof}
}

\begin{proposition}\label{proposition 2}
  Under hypotheses (H1)-(H6), the noise processes $\{\mathbf{u}_{k}\}_{k\geq 0}$ and $\{\mathbf{w}_{k}\}_{k\geq 1}$, involved in (\ref{augmented time-correlated noise eq}) and (\ref{augmented observation eq}), respectively, are sequences of zero-mean mutually uncorrelated random vectors with covariance matrices given by
    $$
      \Sigma_{k}^{\mathbf{w}}=\begin{pmatrix} \Sigma^{w}_{k} & \Sigma^{w^{(12)}}_{k}\\ \Sigma^{w^{(2\hskip .04cm 1)}}_{k}     & \Sigma^{w^{(2)}}_{k} \end{pmatrix}, \ k\geq 1;\quad
      \Sigma_{k}^{\mathbf{u}}=\begin{pmatrix} \Sigma^{u}_{k} & \Sigma^{u^{(12)}}_{k}\\ \Sigma^{u^{(21)}}_{k}    & { \big(\Sigma^{\mathbf{u}}_{k}\big)_{_{22}} } \end{pmatrix}, \ k\geq 0,$$
      in which
      ${\big(\Sigma^{\mathbf{u}}_{k}\big)_{_{22}}}=K_{n_z^2}\left(D_kvec(\mathbf{D}_k\mathbf{F}^T_k)D^T_k
 \otimes \Sigma^u_k\right)K_{n_z^2}+\Sigma^{u^{[2]}}_{k}.$

\vskip.25cm

   Moreover, the stochastic processes $\{ \mathbf{x}_k\}_{k\geq1}$, $\{ \mathbf{u}_k\}_{k\geq 0}$, $\{ \mathbf{w}_k\}_{k\geq 1}$, $\{ {\mathbf{v}}^{\ast}_k\}_{k\geq 1}$  and $\{ {\mathbf{v}}^{\ast\ast}_k\}_{k\geq1}$ are uncorrelated to each other.

\end{proposition}
{
\begin{proof}
The proof is omitted, as it is easily deduced from the model hypotheses.

\end{proof}
}
\begin{remark}

From equations  (\ref{augmented actual measurement eq}) and (\ref{augmented observation eq}), together with the properties established in Propositions \ref{proposition 1} and \ref{proposition 2}, it is immediately deduced that
\begin{equation}\label{Sigma_mathbfzk Sigma_mathbfyk}
\begin{array}{l}
  \Sigma^{\mathbf{z}}_{k}= E[\mathcal{H}_{k}\mathbb{A}_k\mathbb{B}^{T}_k\mathcal{H}^T_{k}]+ \Delta_k  \Upsilon_k^T +\Sigma^{\mathbf{\ddot{v}}}_{k},\quad k\geq 1, \\ \\
  \Sigma^{\mathbf{y}}_{k}=(1-\overline{\lambda}_k)\Sigma^{\mathbf{z}}_{k}+\overline{\lambda}_k \Sigma^{\mathbf{w}}_{k}+\overline{\lambda}_k(1-\overline{\lambda}_k)\mathbf{g}_k\mathbf{g}^T_k,\quad k\geq 1.
\end{array}
\end{equation}
\end{remark}

\section{Quadratic estimation algorithms}
\label{sec:quad_algorithm}
The above statistical properties ensure that the augmented processes involved in equations (\ref{augmented actual measurement eq})-(\ref{augmented observation eq}) all have finite second-order moments. Consequently, the existence of the LS quadratic estimator of $x_k$ based on the original observations (or, equivalently, the LS linear estimator of $x_k$ based on the augmented observations $\mathbf{y}_1, \ldots, \mathbf{y}_L$) is guaranteed. Using an innovation approach, the following recursive algorithm for the quadratic filtering ($L=k$) and fixed-point smoothing ($L=k+N, \ N\geq 1$) estimators is deduced.

\begin{theorem}
  The LS quadratic filtering estimators, $\widehat{x}_{k/k}$, and the error covariance matrices, $\widehat{\Sigma}_{k/k}=E[(x_k-\widehat{x}_{k/k})(x_k-\widehat{x}_{k/k})^T]$, are recursively obtained by
    \begin{equation}\label{filter}
        \widehat{x}_{k/k}=(\mathbb{\breve{A}}_k\ | \ 0)\mathbf{e}_k, \ \ k\geq 1,
    \end{equation}
    \begin{equation}\label{Filtering_error_cov}
        \widehat{\Sigma}_{k/k}=A_kB_k^T-(\mathbb{\breve{A}}_k \ | \ 0)\mathbf{T}_k(\mathbb{\breve{A}}_k \ | \ 0)^T, \ \ k\geq  1,
    \end{equation}
  in which the vectors $\mathbf{e}_k$ and the matrices $\mathbf{T}_k$ satisfy the following recurrence relations
    \begin{equation}\label{recursion_mathbfe_k}
      \mathbf{e}_k=\mathbf{e}_{k-1}+\mathbf{\Psi}_k\Pi^{-1}_k\mu_k,\ \ k\geq 1; \ \ \mathbf{e}_0=0,
    \end{equation}
    \begin{equation}\label{recursion_mathbfT_k}
        \mathbf{T}_k=\mathbf{T}_{k-1}+\mathbf{\Psi}_k\Pi^{-1}_k\mathbf{\Psi}^{T}_k, \ \ k\geq 1; \ \ \mathbf{T}_0=0,
    \end{equation}
  where the innovation $\mu_k$ is calculated by
    \begin{equation}\label{innovation}
      \mu_k= \mathbf{y}_k-(1-\overline{\lambda}_k)\left( \overline{\mathcal{H}}_k\mathbb{A}_k \ | \  \Delta_k\right)\mathbf{e}_{k-1}, \ \ k\geq  1.
    \end{equation}
  The matrices $\mathbf{\Psi}_k$ are given by the following expression
    \begin{equation}\label{mathbfPsi}
      \mathbf{\Psi}_k=(1-\overline{\lambda}_k)\Big( \left( \overline{\mathcal{H}}_k\mathbb{B}_k \ | \  \Upsilon_k\right)-\left( \overline{\mathcal{H}}_k\mathbb{A}_k \ | \  \Delta_k \right)\mathbf{T}_{k-1} \Big)^T , \ \ k\geq 1,
    \end{equation}
   and the innovation covariance matrices, $\Pi_k$, are calculated by
    \begin{equation}\label{innovation covariance matrix}
        \Pi_k=   \Sigma_k^{\mathbf{y}}-(1-\overline{\lambda}_k)^2\left( \overline{\mathcal{H}}_k\mathbb{A}_k \ | \  \Delta_k \right)\mathbf{T}_{k-1}   \left( \overline{\mathcal{H}}_k\mathbb{A}_k \ | \  \Delta_k\right)^T ,  \ \ k\geq  1,
    \end{equation}
  where $\Sigma^{\mathbf{y}}_k$ is obtained by (\ref{Sigma_mathbfzk Sigma_mathbfyk}).

  \vskip.25cm

  At any fixed sampling time $k\geq 1$, by starting from the filter and its error
covariance matrix as initial conditions, the LS quadratic fixed-point smoothers, $\widehat{x}_{k/k+N}$, and
their error covariances, $\widehat{\Sigma}_{k/k+N}$, admit the following recursive relations
    \begin{equation}\label{smoother}
        \widehat{x}_{k/k+N}=\widehat{x}_{k/k+N-1}+ {\mathcal{\breve{S}}}^{x}_{k,k+N}\Pi^{-1}_{k+N}\mu_{k+N}, \ \ N\geq 1,
    \end{equation}
    \begin{equation}\label{Smoothing_error_cov}
        \widehat{\Sigma}_{k/k+N}=\widehat{\Sigma}_{k/k+N-1}-{\mathcal{\breve{S}}}^{x}_{k,k+N}\Pi^{-1}_{k+N}{\mathcal{\breve{S}}}^{xT}_{k,k+N}, \ \ N\geq 1,
    \end{equation}
where
    \begin{equation}\label{mathcal_breveS_smoothing}
      {\mathcal{\breve{S}}}^{x}_{k,k+N}=(1-\overline{\lambda}_{k+N})\left(\big(\mathbb{\breve{B}}_k \ | \  0\big)-\mathbf{M}_{k,k+N-1}\right)
 \left( \overline{\mathcal{H}}_{k+N}\mathbb{A}_{k+N} \ | \  \Delta_{k+N}\right)^T, \ \ N\geq 1,
    \end{equation}
and
     \begin{equation}\label{mathbfM_smoothing}
        \begin{array}{l}
          \mathbf{M}_{k,k+N}=\mathbf{M}_{k,k+N-1}+{\mathcal{\breve{S}}}^{x}_{k,k+N}\Pi^{-1}_{k+N}\mathbf{\Psi}^{T}_{k+N}, \ \ N\geq 1,\\  \mathbf{M}_{k}=  (\mathbb{\breve{A}}_k \ | \ 0)\mathbf{T}_k, \ \ k\geq 1.
        \end{array}
     \end{equation}
\end{theorem}

\begin{proof}
  See Appendix \ref{appendix Proof of Th2}.
\end{proof}

{

\subsection{Computational procedure}
\vskip .15cm
Next, the computational procedure of the proposed  quadratic filtering and fixed-point smoothing algorithms  is summarized.

\begin{itemize} [leftmargin=\dimexpr 24pt-.14in]
  \item [{\em 1)}]

  {\em Covariance matrices of the augmented processes.}

\begin{itemize}
  \item [\em  1a)] The covariance matrices $\Sigma^\mathbf{x}_{k,s}$, $\Sigma^{{\mathbf{v}}^{\ast}}_{k,s}$, $\Sigma^{{\mathbf{v}}^{\ast\ast}}_{k}$, $\Sigma^{\mathbf{w}}_{k}$ and $\Sigma^{\mathbf{u}}_{k}$ of the augmented signal and noise processes  are obtained by using the expressions established in Propositions 1 and 2.

   \item [\em  1b)] From the matrices obtained in {\em  1a)}, the covariance matrices $\Sigma^\mathbf{z}_{k}$ and $\Sigma^\mathbf{y}_{k}$ are  computed by (\ref{Sigma_mathbfzk Sigma_mathbfyk}).
  \end{itemize}

All these covariance matrices only depend on the system model information, so they can be calculated offline, before the observations are available.

\vskip .1cm
  \item [{\em 2)}]

{\em   LS quadratic filtering recursive algorithm.}
At the sampling time $k$,  starting with the prior knowledge
of the $(k-1)$-th iteration   (consequently,  $\mathbf{\Psi}_{k-1}$, $\Pi_{k-1}$, $\mathbf{T}_{k-1}$ $\mu_{k-1}$ and $\mathbf{e}_{k-1}$ are known), the proposed quadratic filtering algorithm operates as follows:

\vskip .1cm

\begin{itemize}
  \item [\em  2a)] {\em Filtering error covariance matrices.} Compute $\mathbf{\Psi}_{k}$ by (\ref{mathbfPsi})
   and, from it and  with $\Sigma^\mathbf{y}_{k}$ obtained in {\em  1a)}, the innovation covariance
  matrix $\Pi_{k}$ is provided by (\ref{innovation covariance matrix}).
  Then  $\mathbf{T}_{k}$  is obtained by (\ref{recursion_mathbfT_k})
   and, from it, the filtering error covariance matrices,
$\widehat{\Sigma}_{k/k}$, are obtained by (\ref{Filtering_error_cov}).
It should be noted that theses matrices do not depend on the measurements, thus providing a measure of the filter performance even before we get any observed data.

\vskip .1cm

  \item [\em  2b)] {\em Quadratic filtering estimators.} When the new measurement $\mathbf{y}_k$ is available, the innovation $\mu_k$ is computed by
 expression (\ref{innovation}),
  and, from it, $\mathbf{e}_k$ is obtained by (\ref{recursion_mathbfe_k}).
  Then, the quadratic filtering estimators, $\widehat{x}_{k/k}$
 are computed by (\ref{filter}).
\end{itemize}

\vskip .1cm

  \item [{\em 3)}]
 {\em   LS quadratic fixed-point smoothing recursive algorithm.} At any fixed sampling time $k\geq 1$, once the filter, $\widehat{{x}}_{k/k}$, and the filtering error covariance matrix,  $\widehat{\Sigma}_{k/k}$ are available, the proposed quadratic smoothing estimators  and the corresponding error covariance matrix are obtained as follows:

 For $N=k+1,k+2,\ldots,$ compute the  matrices $\mathbf{M}_{k,k+N-1}$ using  (\ref{mathbfM_smoothing})
  and, from these matrices, $ {\mathcal{\breve{S}}}^{x}_{k,k+N}$ is derived by (\ref{mathcal_breveS_smoothing});
  then, the smoothers $\widehat{x}_{k/k+N}$ and their error covariance
 matrices $\widehat{\Sigma}_{k/k+N}$ are obtained from (\ref{smoother})
 and (\ref{Smoothing_error_cov}),
 respectively.

\end{itemize}

}

\section{Simulation study}
\label{sec:simul_study}

In this section, a simulation numerical example is considered to analyze the implementation and performance of the proposed quadratic filtering and  fixed-point smoothing algorithm.

\vskip .2cm
\noindent {\em AR(1) scalar signal.}
Consider   a scalar signal process $\{x_k\}_{ k\geq1}$  generated by the following first-order autoregressive model: $$ x_{k}=0.95x_{k-1}+\varepsilon_{k-1}, \ \
\ k \geq 1,$$ where the  initial signal  $x_0$ is a zero-mean Gaussian variable with variance $\Sigma^x_{0}=0.1$,
and $\{\varepsilon_k\}_{ k\geq0}$ is a zero-mean white Gaussian noise with variance $\Sigma^{\varepsilon}_{k} =0.1, \ \forall k\geq 0$.

Assuming that $x_0$ and the sequence $\{\varepsilon_k\}_{ k\geq0}$ are mutually independent and taking into account that the third and fourth-order moments of a zero-mean Gaussian variable with variance $\sigma^2$ are 0 and $3\sigma^4$, respectively,  the  covariance and cross-covariance functions of this signal and their second-order powers  are given by
$$\Sigma^x_{k,s}= 0.95^{k-s}\Sigma_s^x, \ \ \ \Sigma^{x^{(2)}}_{k,s}= 0.95^{2(k-s)}\Sigma^{x^{(2)}}_{s}, \ s\leq k; \ \ \ \ \Sigma^{x^{(12)}}_{k,s}=0, \ \forall k,s,$$
where the functions $\Sigma^x_s$ and $\Sigma^{x^{(2)}}_{s}$ are recursively obtained by
$$\Sigma^x_s=0.9025\,\Sigma^x_{s-1}+ 0.1, \ s\geq 1,$$
 $$\Sigma^{x^{(2)}}_s=0.8145\,\Sigma^{x^{(2)}}_{s-1}+ 0.361\,\Sigma^x_{s-1}+0.02, \ s\geq 1.$$
According to assumption {\em (H1)}, it is clear that these covariance functions can be expressed in a separable form  defining, for example, the following functions:
$$\begin{array}{l}A_k= 0.95^k, \ B_k= 0.95^{-k}\Sigma^x_k; \ \ \ \breve{A}_k= 0.95^{2k}, \ \breve{B}_k= 0.95^{-2k}\Sigma^{x^{(2)}}_k;\\
A_{1,k}= B_{1,k}= 0; \ \ \ A_{2,k}= B_{2,k}= 0.\end{array}$$
\vskip .2cm
\noindent {\em Actual measurements.}
Assume that the real measurements of the signal, $z_k$, are described by the  model (\ref{actual_measurements}) with  the following parameters:
\begin{itemize}[leftmargin=\dimexpr 26pt-.2in]
  \item[$\circ$] $H_k=0.9\theta_k$,  in which   $\{\theta_k\}_{k\geq 1}$ is a sequence of independent  identically distributed Bernoulli random  variables with probability $P(\theta_k=1)=\overline{\theta}$. These variables model whether the signal is present ($\theta_k=1$) or not ($\theta_k=0$) in the actual measurements and, therefore, $\overline{\theta}$ is the probability that the observations contain the signal to be estimated.

  \item [$\circ$]  The noise process  $\{v_{k}\}_{k\geq 0}$ is generated by (\ref{time-correlated_noise}) where  $D_k=0.75$,  $\{u_k\}_{ k\geq0}$ is a zero-mean white Gaussian noise with $\Sigma^{u}_{k} =0.01, \ \forall k\geq 0$, and  $v_0$ is a zero-mean Gaussian variable with $\Sigma^v_{0}=0.1$; hence,
      $$
\Sigma^{v^{(12)}}_0=\Sigma^{u^{(12)}}_k=0, \ \ \Sigma^{v^{(2)}}_0= 0.02, \ \
\Sigma^{u^{(2)}}_k= 0.0002.
$$
\end{itemize}

According to the theoretical model, let us suppose that the measurements
are subject to deception attacks and the signal injected by the adversaries is given by (\ref{deception_attack}). The false
data injection attack noise $\{w_{k}\}_{k\geq 1}$ is a white non-Gaussian  sequence
with  distribution
\[
P(w_k=-8)=1/8, \ \ \ P(w_k=8/7)
=7/8, \ \ \ \ \forall k\geq 1;
\]
 hence,
$$
E[w_k]=0, \ \ \Sigma^{w}_{k}=9.1429, \ \ \Sigma^{w^{(12)}}_k=-62.6939, \ \
\Sigma^{w^{(2)}}_k= 429.9009.
$$

\noindent{\em Available observations.} Finally, again in line with our theoretical study, we suppose that the available observations for the estimation are given by (\ref{available_observ}), where the white sequence of Bernoulli random variables  $\{\lambda_k\}_{ k\geq 1}$, modeling whether the deception attacks actually succeed or
not, are identically distributed with probabilities $P(\lambda_k=1)=\overline{\lambda}.$

\vskip .2cm
Our goal with this example is threefold. First, we aim at showing the feasibility and effectiveness of
the proposed quadratic estimators, illustrating their performance and the superiority of the quadratic estimators over the linear ones (the linear filtering and fixed point smoothing algorithms are given in Appendix \ref{appendix linear estimation algorithm}).
Second, we intend to show how the probability  $\overline{\theta}$ that the signal is present in the  actual measurements influence the
performance of the estimators.
Third, we attempt to show the effect of the successful deception attack probability $\overline{\lambda}$ over the
performance of the estimators.

For this purpose, a MATLAB program has been developed to obtain the linear and quadratic estimators and, in order to quantify the estimation accuracy, the corresponding estimation error variances were calculated for different values of the probabilities $\overline{\theta}$ and $\overline{\lambda}$.

\vskip .25cm

\begin{figure}[H]
\centering
\includegraphics[width=13 cm]{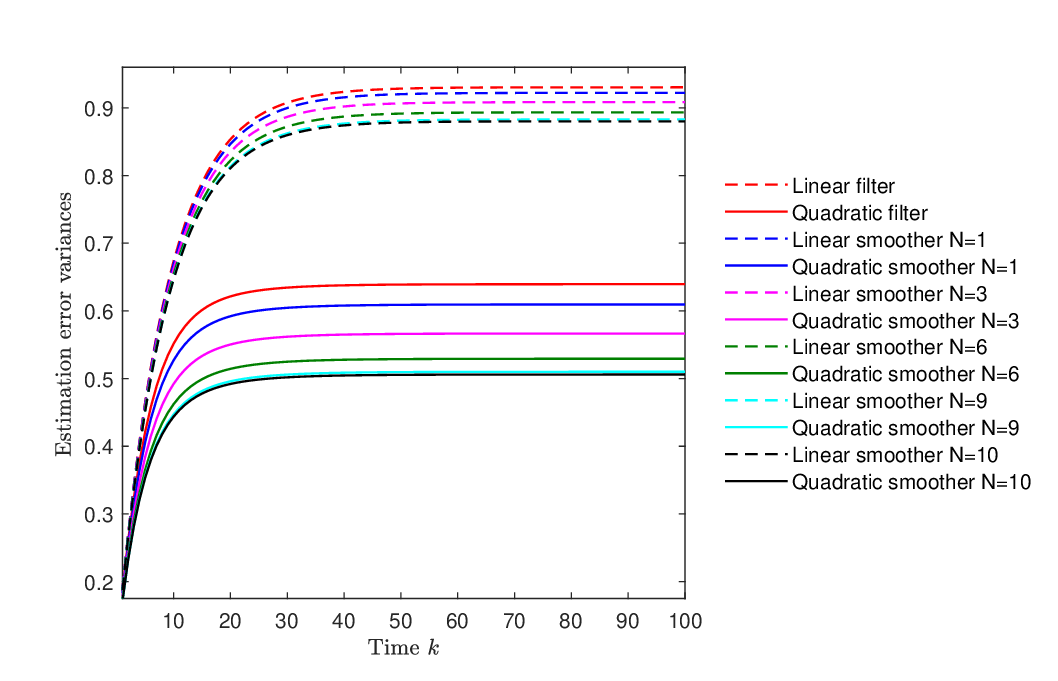}
\caption{Estimation error variance comparison of the linear and quadratic filtering and smoothing estimators when $\overline{\theta}=\overline{\lambda}=0.5$.}\label{fig1}
\end{figure}

\noindent{\em Performance of the quadratic filtering and fixed-point smoothing estimators.}
Considering  the same fixed value 0.5 for the probabilities  $\overline{\theta}$  and $\overline{\lambda}$, the error variances of the linear and quadratic estimators are calculated to compare the performance of both  filtering and fixed-point smoothing estimators.
The results of this comparison  are displayed in Figure~\ref{fig1}, which shows, on the one hand, that the  quadratic estimators present lower error variances than the linear ones, thus confirming the superiority of the former over the latter.
On the other hand, it is gathered that, for both linear and quadratic estimators, the smoothing error variances are less than the corresponding filtering ones and, also,
 that as the number of available observations increases, the fixed-point smoothers become more accurate. {
Furthermore, it is
observed that the values of the fixed-point smoothing
error variance decrease with increasing $N$, although this
decrease becomes almost negligible for $N \geq 9$.}

\begin{figure}[H]
\centering
\vskip-0cm\includegraphics[width=13 cm]{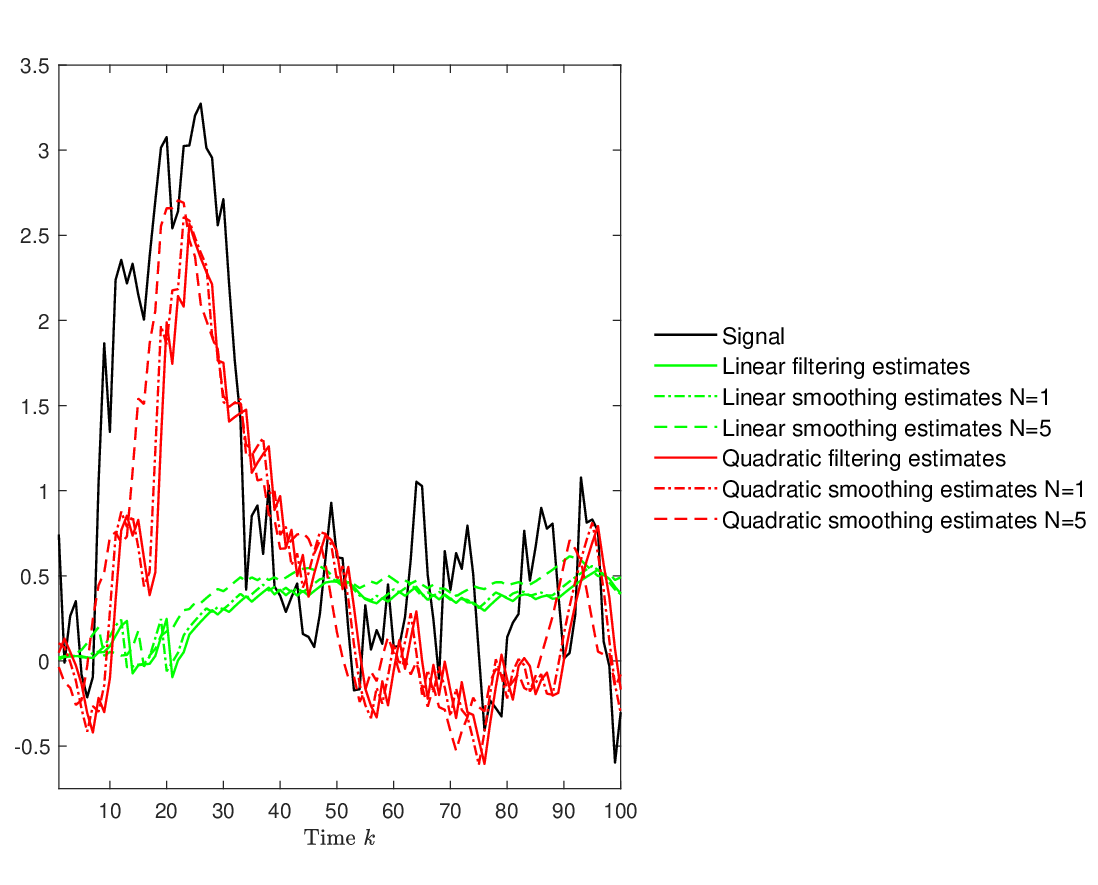}
 \caption{Simulated signal, linear and quadratic filtering { and fixed-point smoothing } estimates when $\overline{\theta}=\overline{\lambda}=0.5$.}\label{fig2}
  \end{figure}

Figure~\ref{fig2}  displays a simulated signal trajectory and their corresponding linear and quadratic filtering { and smoothing} estimates. Agreeing with the comments made about Figure~\ref{fig1}, it is observed that the quadratic filtering  { and smoothing} estimates track the signal evolution better than the linear ones.
{ It is also noticed that the accuracy of the
quadratic  smoothing estimate is higher than that of the quadratic filtering estimate.}
\vskip-.5cm

\begin{figure}[H]
\centering
\includegraphics[width=10 cm]{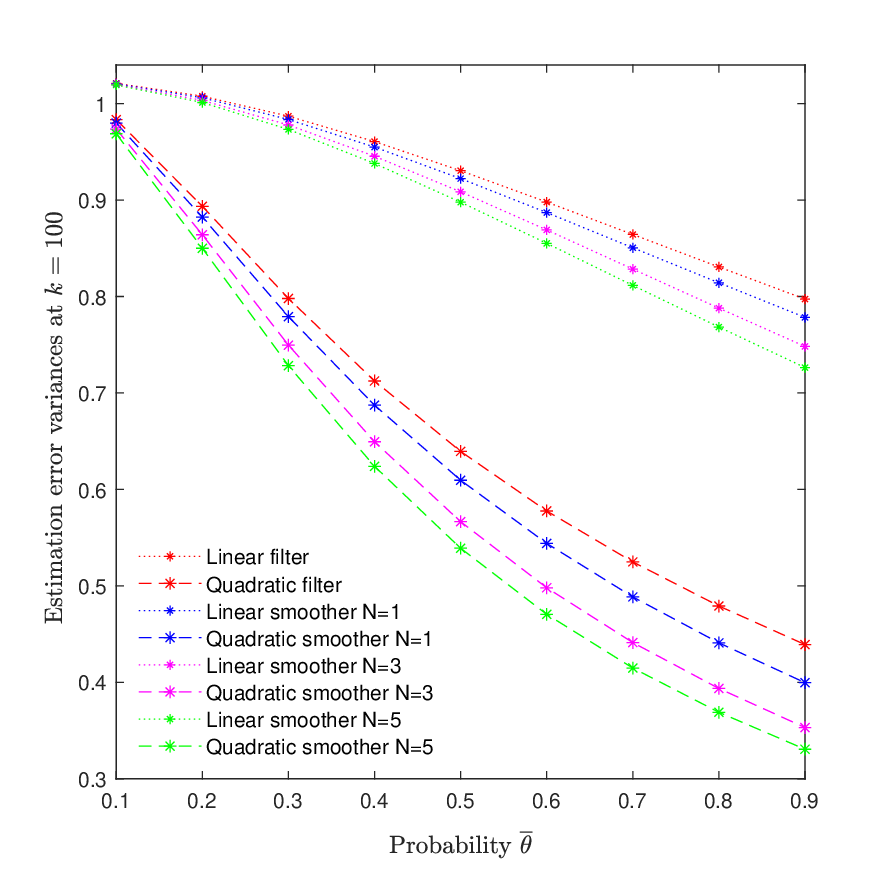}
\caption{Linear and quadratic filtering and smoothing error variances versus $\overline{\theta}$, when $\overline{\lambda}=0.5$.}\label{fig3}
\end{figure}

\noindent{\em Influence of the probability  $\overline{\theta}$.}
Assuming, as in the above figures, that the attack  probability is $\overline{\lambda}=0.5$, now we compare the performance of the estimators  considering
different values of the  probability, $\overline{\theta}$, that the signal is present in the  actual measurements.
Since, from $k=50$ onwards the estimation error variances show a similar behaviour, only the variances at a specific iteration $k=100$ are
considered.
 To illustrate the influence of the probability  $\overline{\theta}$,  Figure~\ref{fig3} depicts the comparative results between the filtering and smoothing error variances for both linear and quadratic estimators, considering several values of the probability $\overline{\theta}$ (namely, $\overline{\theta}\!= 0.1$ to  $0.9$).
This figure shows that $\overline{\theta}$ --the probability that the observations contain the signal-- or, equivalently,  the probability $1- \overline{\theta}$ that the signal is missing in the  actual measurements,  indeed influence the performance of the estimators.
Actually, as expected, both linear and quadratic estimation error variances decrease
as  $\overline{\theta}$ increases and, consequently, the filtering and smoothing estimators perform better when the probability  that the signal is missing in the  actual measurements, $1- \overline{\theta}$, decreases.
As in Figure~\ref{fig1}, this figure also shows that, for all the values of $\overline{\theta}$, the quadratic estimation error variances are smaller than the linear ones; besides, it is observed that the smoothing estimation error variances, for both the linear and quadratic estimators, are lower than those of the filters, and that the smoother  performs better as the number of available observations increases.
It is also inferred  that,  as the values of the probability $\overline{\theta}$ increase, a higher reduction in the estimation error variances is yielded by the quadratic filtering and smoothing estimators over the linear ones.
\vskip-.5cm

\vskip 0.25cm

\begin{figure}[H]
\centering
  \includegraphics[height=10cm]{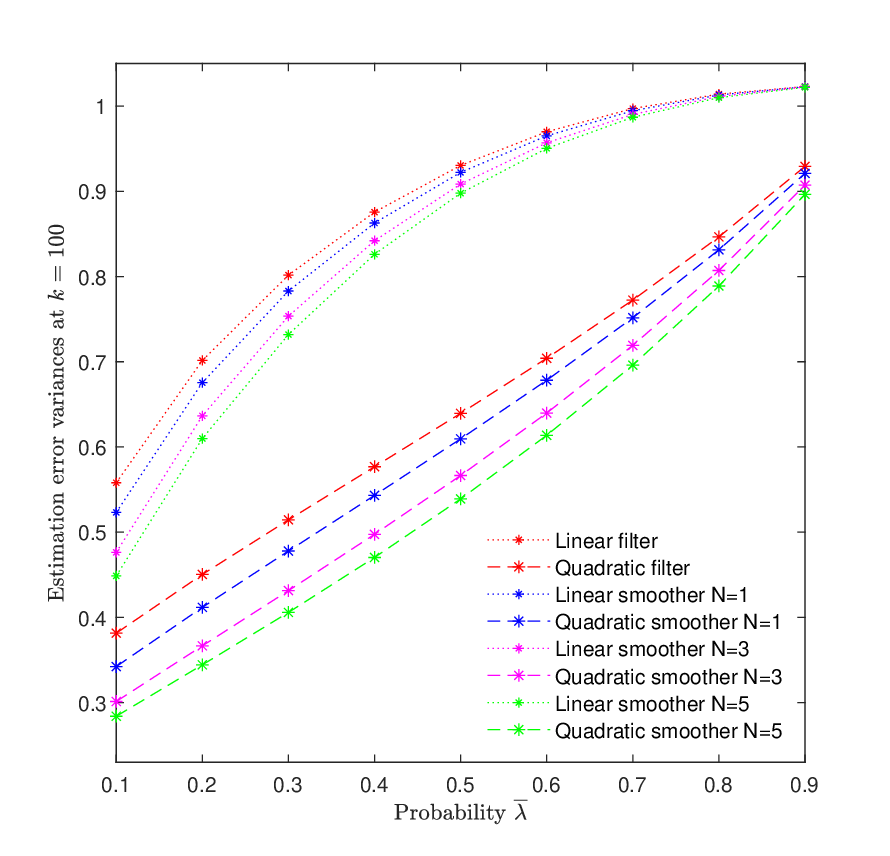}
\caption{Linear and quadratic filtering and smoothing error variances versus $\overline{\lambda}$, when $\overline{\theta}=0.5$.}\label{fig4}
\end{figure}

\noindent{\em Effect of the attack probability $\overline{\lambda}$.}
Assuming again, as in Figure~\ref{fig1}, that $\overline{\theta}=0.5$, we examine the impact of the deception attacks on the estimation accuracy. More precisely, we compare the performance of the estimators  considering
several values of the successful deception attack probability $\overline{\lambda}\!= 0.1$ to  $0.9$. As $\overline{\lambda}$ increases, the number of successful  attack is expected to be greater and, consequently, a higher number of available measurements used for estimation  will be only noise; so, worse estimations will be obtained and, hence, the error variances are expected to be higher. Figure~\ref{fig4} confirms this fact, showing that
 the filtering and smoothing error variances at $k=100$, of both linear and quadratic estimators,  become smaller as the successful deception attack probability $\overline{\lambda}$ decreases. This figure also shows that,  in the case of quadratic estimators, similar increments in the values of the probability $\overline{\lambda}$ produce essentially the same increase of the estimation error variances. However, in the linear estimation problem, such increase is more significant for small values of $\overline{\lambda}$.

\section{Conclusion}
\label{sec:conclusion}

Recursive algorithms for the LS quadratic filtering and fixed-point
smoothing estimation problems are proposed from measurements perturbed
by random parameter matrices, time-correlated additive noises and random deception
attacks. Unlike most studies on quadratic estimation, in which the linear estimator of the augmented signal is calculated and, from it, the estimator of the original signal is extracted, we deal with the direct estimation of the original signal based on the augmented observations. Some numerical results are used to examine the accuracy of the quadratic estimators, which reveal that the proposed estimators outperform the linear ones and illustrate how the theoretical system model under consideration covers the missing measurements phenomenon as a specific example. In addition, the effect of missing measurement and deception attack success probabilities
 on the estimation accuracy are
analyzed in the context of the numerical simulation study.

{
Future research topics would include extending the proposed framework to deal
with more sophisticated attack models, such as the important-data-based attack model used in \cite{Wang2023}.}
{ It would also be interesting to consider the distributed estimation problem in the scenario of networked systems whose sensor nodes are spatially distributed and connected according to a predetermined topology (see \cite{Liu2023}).}

\section*{Funding}

This research was suported by the “Ministerio de Ciencia e Innovación, Agencia Estatal de Investigación” of Spain
and the European Regional Development Fund [grant number PID2021-124486NB-I00].

\appendix

\section{Proof of Theorem 1}
 \label{appendix Proof of Th1}

\begin{itemize}[leftmargin=\dimexpr 26pt-.1in]
  \item[\em(a)]
   Taking into account that $E[H_{k}x_{k}\otimes v_{k}]=0$, it is clear that $\overline{\mathcal{V}}^*_{k}=0$; then, taking expectations in (\ref{augmented actual observ}), we have that
$\overline{\mathcal{Z}}_{k}=\overline{\mathcal{H}}_{k}\overline{\mathcal{X}}_{k}+\overline{\mathcal{V}}_{k},\ k\geq 1.$ Hence, again from (\ref{augmented actual observ}), by adding and subtracting $\mathcal{H}_{k}\overline{\mathcal{X}}_{k}$, we obtain
$$\mathbf{z}_k=\mathcal{H}_{k}\mathbf{x}_k+(\mathcal{H}_{k}-\overline{\mathcal{H}}_{k})\overline{\mathcal{X}}_{k}
+\mathbf{v}_k+\mathcal{V}^*_{k}-\mathbf{\nu}_k+\mathbf{\nu}_k, \ \ k\geq 1,
$$
where $ \mathbf{\nu}_k=\left(
\begin{array}
[c]{c}
0  \\K_{n_z^{2}}\big(\overline{H}_kx_k)
 \otimes v_k\big)
\end{array}
\right)$.

Clearly, ${\mathbf{v}}^{\ast}_k=\mathbf{v}_{k}+\mathbf{\nu}_k$ and, taking into account that
$$(\mathcal{H}_{k}-\overline{\mathcal{H}}_{k})\overline{\mathcal{X}}_{k}=\left(
\begin{array}
[c]{c}
0  \\ \big({H}^{[2]}_k-\overline{H}^{[2]}_k\big) vec(A_kB^T_k)
\end{array}
\right)$$ it is obvious that ${\mathbf{v}}^{\ast\ast}_k=(\mathcal{H}_{k}-\overline{\mathcal{H}}_{k})\overline{\mathcal{X}}_{k}+\mathcal{V}^*_{k}-\mathbf{\nu}_k$; so, equation (\ref{augmented actual measurement eq}) is directly obtained.

Finally, taking expectations in  (\ref{time-correlated augmented noise}), we have
$ \overline{\mathcal{V}}_{k}=\mathcal{D}_{k-1}\overline{\mathcal{V}}_{k-1}+\overline{\mathcal{U}}_{k-1},\ k\geq 1,$ and equation (\ref{augmented time-correlated noise eq})  is straightforward.

  \item[\em(b)] Taking expectations in equation (\ref{augmented observation}), it is clear that
  $\overline{\mathcal{Y}}_{k}=(1-\overline{\lambda}_k)\overline{\mathcal{Z}}_{k}+\overline{\lambda}_k\overline{\mathcal{W}}_{k},\ k\geq 1,$ and hence
$$\mathbf{y}_k=(1-\lambda_k)\mathcal{Z}_{k}
-(1-\overline{\lambda}_k)\overline{\mathcal{Z}}_{k}+\lambda_k\mathcal{W}_{k}-
\overline{\lambda}_k\overline{\mathcal{W}}_{k},\ k\geq 1.$$
Now, by adding and subtracting both $\lambda_k\overline{\mathcal{W}}_{k}$ and $(1-\lambda_k)\overline{\mathcal{Z}}_{k}$, we obtain
$$\mathbf{y}_k=(1-\lambda_k)\mathbf{z}_{k}+\lambda_k \mathbf{w}_{k}-(\lambda_k-\overline{\lambda}_k)\big(\overline{\mathcal{Z}}_{k}-\overline{\mathcal{W}}_{k}\big),\ k\geq 1,$$
and denoting
$$\mathbf{g}_k=\overline{\mathcal{Z}}_{k}-\overline{\mathcal{W}}_{k}
=\displaystyle\binom{0}{\overline{H}_{k}^{[2]}vec(A_kB^T_k)+vec(\mathbf{D}_k\mathbf{F}^T_k)}-\displaystyle\binom{0}{vec(\Sigma^w_k)},$$ expression (\ref{augmented observation eq}) holds.

  \end{itemize}

{
\section{Proof of Proposition 1}
 \label{appendix Proof of Prop1}

\begin{itemize}[leftmargin=\dimexpr 26pt-.1in]

  \item[\em(a)] Expression (\ref{Covx}) for $\Sigma^\mathbf{x}_{k,s}$ is obtained taking into account that, from
hypothesis (H1),
  $$\Sigma^\mathbf{x}_{k,s}= \left(
\begin{array}
[c]{cc}
A_kB^T_s     & A_{1,k}B^T_{1,s}   \\
A_{2,k}B^T_{2,s}      & \breve{A}_k\breve{B}^T_s
\end{array}
\right),\  s\leq k.$$
The  factorization (\ref{Ex}) for the expectations $E[x_k\mathbf{x}^{T}_s]$ is immediately deduced using that, again from hypothesis (H1),
  $$E[x_k\mathbf{x}^{T}_s]=\left\{
\begin{array}
[c]{l}
 \left(A_kB^T_s  \ | \  A_{1,k}B^T_{1,s} \right),\ \  s\leq k,  \\
 \left(B_kA^T_s  \ | \  B_{2,k}B^T_{2,s} \right),\  \ k\leq s.
\end{array}
\right.
$$

   \item[\em(b)]
 We write ${\mathbf{v}}^{\ast}_k=\mathbf{v}_{k}+\mathbf{\nu}_k$, where $ \mathbf{\nu}_k=\left(
\begin{array}
[c]{c}
0  \\K_{n_z^{2}}\big(\overline{H}_kx_k)
 \otimes v_k\big)
\end{array}
\right)$.

Since, from the independence hypotheses on the model, the signal $x_k$ and the noise $v_s$ are  uncorrelated, it is clear that $\mathbf{v}_{k}$ and $\mathbf{\nu}_s$ are also uncorrelated and, hence, $\Sigma_{k,s}^{\mathbf{v}^\ast}=\Sigma_{k,s}^{\mathbf{v}}+\Sigma_{k,s}^{\nu}$. Next, these covariance matrices are obtained:

\begin{itemize}[leftmargin=\dimexpr 26pt-.2in]
  \item From (\ref{augmented time-correlated noise eq}) and assumptions (H3) and (H4), the covariance function $\Sigma_{k,s}^{\mathbf{v}}$  is factorized in a separable form, $\Sigma_{k,s}^{\mathbf{v}} = \mathbb{D}_k \mathbb{F}_s^T$, $s\leq k$,  in which
      $\mathbb{D}_k = \mathcal{D}_{k,0} $ and
      $\mathbb{F}_s^T = \mathcal{D}_{s,0}^{-1} \Sigma_s^{\mathbf{v}}$,
       with $\mathcal{D}_{k,0}=\mathcal{D}_{k-1} \cdots \mathcal{D}_0$. Again, from (\ref{augmented time-correlated noise eq}) and the model hypotheses,  starting from $\Sigma_0^{\mathbf{v}}$, the matrices  $\Sigma_s^{\mathbf{v}}$ are recursively computed by
$\Sigma_s^{\mathbf{v}} = D_{s-1} \Sigma_{s-1}^{\mathbf{v}} D_{s-1}^T + \Sigma^{\mathbf{u}}_{s-1},\  s\geq 1.$

  \item Taking into account that, from (H1), $E[x_kx^T_s]=A_kB^T_s, \ s\leq k,$ and, from Remark 5, $E[v_kv^T_s]=\mathbf{D}_k\mathbf{F}^T_s,\ s\leq k,$ the Kronecker product properties lead us to

      $E\big[\left(\overline{H}_kx_k \otimes v_k\right)\!\left(\overline{H}_sx_s \otimes v_s\right)^T\big]
\!=\!\big(\overline{H}_kA_kB^T_s\overline{H}^T_s\big) \otimes \mathbf{D}_k \mathbf{F}_s^T$

$\hskip 5.5cm=\!\big((\overline{H}_kA_k)\otimes \mathbf{D}_k\big)\big( (\overline{H}_sB_s) \otimes  \mathbf{F}_s\big)^T\!\!, \ s\leq k.$

So,  it is clear that $\Sigma_{k,s}^{\nu}$ is factorized  as $\Sigma_{k,s}^{\nu} = \mathbb{C}_k \mathbb{E}_s^T$, $s\leq k$.
\end{itemize}

The above two items guarantee that
$\Sigma_{k,s}^{\mathbf{v}^\ast}=\mathbb{D}_k \mathbb{F}_s^T+\mathbb{C}_k \mathbb{E}_s^T, \ s \leq k,$  so  the  factorization (\ref{Cov-v-ast}) for $\Sigma_{k,s}^{\mathbf{v}^\ast}$ is immediately obtained.

    \item[\em(c)] Clearly, $E\big[\mathbf{v}^{\ast\ast}_{k}\big]=0$. Next,
    using again the independence hypotheses on the model and  the Kronecker product properties, we have:

\begin{itemize} [leftmargin=\dimexpr 26pt-.2in]

  \item[$\bullet$]  $E\big[\big(\widetilde{H}^{[2]}_kvec(A_kB^T_k)\big)\big(\widetilde{H}^{[2]}_svec(A_sB^T_s)\big)^T \big]$

\hskip 5cm   $=E\big[\widetilde{H}^{[2]}_kvec(A_kB^T_k)vec^T\!(A_kB^T_k) \widetilde{H}^{[2]T}_k\big] \delta_{k,s}.$

  \item[$\bullet$]  $E\big[\big(\widetilde{H}^{[2]}_kvec(A_kB^T_k)\big)\big(\widetilde{H}_sx_s \otimes v_s\big)^T\big] =0.$

  \item[$\bullet$] $E\Big[\big((\widetilde{H}_kx_k) \otimes v_k\big)\big(\widetilde{H}_sx_s \otimes v_s\big)^T\Big]=E\big[\widetilde{H}_kE[x_kx^T_k]\widetilde{H}_k\big]\delta_{k,s} \otimes E[v_kv^T_s]\Big)$

\hskip 5.75cm $   =\Big(E\big[\widetilde{H}_kA_kB^T_k\widetilde{H}_k\big] \otimes \mathbf{D}_k \mathbf{F}_k^T\Big)\delta_{k,s}.$

\end{itemize}

From the above items,  we conclude that, for $k\neq s$, $E\big[\mathbf{v}^{\ast\ast}_{k}\mathbf{v}^{\ast\ast T}_{s}\big]=0$; so, the uncorrelation of the vectors is proven. Also, expression (\ref{22-block}) for $\big(\Sigma^{{\mathbf{v}}^{\ast\ast}}_{k}\big)_{_{22}}$, the $(2, 2)$-block of the matrix $\Sigma^{{\mathbf{v}}^{\ast\ast}}_{k}$,
 is straightforward.

  \end{itemize}
}

\section{Proof of Theorem 2}
 \label{appendix Proof of Th2}

Since the quadratic estimator, $\widehat{x}_{k/L}$,  of the signal $x_k$ based on the observations $y_1, \ldots, y_L$ is equal to the linear estimator of $x_k$ based on the augmented observations $\mathbf{y}_1, \ldots, \mathbf{y}_L$, according to the innovation approach it  can be expressed as a linear combination of the innovations $\mu_1, \ldots, \mu_L$; namely,
\begin{equation}\label{general expression estimators}
  \widehat{x}_{k/L}=\displaystyle{\sum_{h=1}^{L}}
 {\mathcal{\breve{S}}}^{x}_{k,h}\Pi^{-1}_h\mu_h, \  \ k, L \geq 1,
\end{equation}
where ${\mathcal{\breve{S}}}^{x}_{k,h}=E[x_k\mu^{T}_h]$, \  $\mu_h = \mathbf{y}_h- \widehat{\mathbf{y}}_{h/h-1}$ \ and \
$\Pi_h=E[\mu_h \mu^T_h]$. To begin with, we are going to derive a proper expression for the one-stage observation predictors $\widehat{\mathbf{y}}_{h/h-1}$, that allows us to calculate the innovations $\mu_h$ and, from them, the coefficients $ {\mathcal{\breve{S}}}^{x}_{k,h}$ and the innovation covariance matrices $\Pi_h$ involved in the general expression of the estimators (\ref{general expression estimators}).

Taking into account equations (\ref{augmented actual measurement eq}) and (\ref{augmented observation eq}), together with the incorrelation properties established in Proposition \ref{proposition 2} and the Orthogonal Projection Lemma (OPL), it is clear that
\begin{equation}\label{observ_predictor}
  \widehat{\mathbf{y}}_{k/k-1}= (1-\overline{\lambda}_k)\big(\overline{\mathcal{H}}_k \widehat{\mathbf{x}}_{k/k-1}
+ \widehat{{\mathbf{v}}}^{\ast}_{k/k-1}\big), \ \ k\geq 1.
\end{equation}
So, the one-stage predictor of both the augmented signal, $\widehat{\mathbf{x}}_{k/k-1}$, and the augmented noise, $\widehat{\mathbf{v}}^{\ast}_{k/k-1}$,  must be calculated. Similarly to (\ref{general expression estimators}), denoting ${\mathcal{S}}^{\mathbf{x}}_{k,h}=E[\mathbf{x}_k\mu^{T}_h]$ and ${\mathcal{S}}^{{\mathbf{v}}^{\ast}}_{k,h}=E[{\mathbf{v}}^{\ast}_k\mu^{T}_h]$, these estimators can be obtained by the following general expressions:
\begin{equation}\label{general expression estimators mathbfx mathbfdotv}
  \widehat{\mathbf{x}}_{k/L}=\displaystyle{\sum_{h=1}^{L}}
 {\mathcal{S}}^{\mathbf{x}}_{k,h}\Pi^{-1}_h\mu_h, \  \  \ \  \widehat{{\mathbf{v}}}^{\ast}_{k/L}=\displaystyle{\sum_{h=1}^{L}}{\mathcal{S}}^{{\mathbf{v}}^{\ast}}_{k,h}\Pi^{-1}_h\mu_h; \  \  \ \ k, L \geq 1.
\end{equation}

\vskip.25cm

\noindent $\bullet$ \emph{One-stage predictor and filter of the signal: $\widehat{x}_{k/s}, \ s\leq k$.}

\vskip.25cm

From (\ref{observ_predictor}), it is straightforward to see that
$$
{\mathcal{\breve{S}}}^{x}_{k,h}= E\big[x_k \mathbf{y}_{h}^T\big]-(1-\overline{\lambda}_h)\Big(E\big[
 x_k\widehat{\mathbf{x}}^{T}_{h/h-1}\big]\overline{\mathcal{H}}_h^T + E\big[x_k\widehat{{\mathbf{v}}}^{\ast T}_{h/h-1}\big]
 \Big), \ \ 1\leq h \leq k.
$$
Now, taking into account Proposition \ref{proposition 1}(a) and Proposition \ref{proposition 2}, it is clear that $$E\big[x_k\mathbf{y}^{T}_{h}\big]=(1-\overline{\lambda}_h)\mathbb{\breve{A}}_k\mathbb{B}_h^{T}\overline{\mathcal{H}}_h^T, \ \ 1\leq h \leq k,$$ and, using (\ref{general expression estimators mathbfx mathbfdotv}), we can write
$$
  E\big[ x_k\widehat{\mathbf{x}}^{T}_{h/h-1}\big] =\displaystyle{\sum_{j=1}^{h-1}}{\mathcal {\breve{S}}}^{x}_{k,j}\Pi^{-1}_j {\mathcal S}_{h,j}^{{\mathbf{x}}T}, \ \ \ \
  E\big[ x_k\widehat{{\mathbf{v}}}^{\ast T}_{h/h-1}\big] =\displaystyle{\sum_{j=1}^{h-1}}{\mathcal {\breve{S}}}^{x}_{k,j}\Pi^{-1}_j {\mathcal S}_{h,j}^{\mathbf{v}^{\ast }T}, \ \ \ \ h\geq 2.
$$
Consequently,
$$
{\mathcal {\breve{S}}}^{x}_{k,h}\hskip -0.05cm =
(1-\overline{\lambda}_h)\Big[ \ \mathbb{\breve{A}}_k\mathbb{B}_h^T \overline{\mathcal{H}}_h^T -(1-\delta_{h,1})\displaystyle{\sum_{j=1}^{h-1}}
{\mathcal{\breve{S}}}^{x}_{k,j}\Pi^{-1}_j \big( \overline{\mathcal{H}}_h{\mathcal S}^{\mathbf{x}}_{h,j} + {\mathcal S}_{h,j}^{\mathbf{v}^{\ast}} \big)^T \Big],\  \ h\geq 1.
  $$
  Hence, if we define
\begin{equation}\label{brevePsimathbfx}
\Psi^\mathbf{x}_h=(1-\overline{\lambda}_h)\Big[\ \mathbb{B}_h^T \overline{\mathcal{H}}_h^T -(1-\delta_{h,1})\displaystyle{\sum_{j=1}^{h-1}}\Psi^\mathbf{x}_{j}\Pi^{-1}_j \big( \overline{\mathcal{H}}_h{\mathcal S}^{\mathbf{x}}_{h,j}+ {\mathcal S}_{h,j}^{\mathbf{v}^{\ast}} \big)^T \Big],\  \ h\geq 1,
\end{equation}
we can write
$
{\mathcal{\breve{S}}}^{x}_{k,h}= \mathbb{\breve{A}}_k\Psi^\mathbf{x}_h, \ h\leq k.
$
So, denoting
\begin{equation}\label{ex}
e^\mathbf{x}_k = \displaystyle{\sum_{h=1}^{k}}\Psi^\mathbf{x}_h\Pi^{-1}_h\mu_h,\ \ k\geq 1; \ \ \ e^\mathbf{x}_0 = 0,
\end{equation}
and using (\ref{general expression estimators}), we conclude that
\begin{equation}\label{signal_pred_and_filter}
\widehat{x}_{k/s}=\mathbb{\breve{A}}_k e^\mathbf{x}_s, \ \ 1\leq s\leq k.
\end{equation}

\vskip.25cm

\noindent $\bullet$  \emph{One-stage predictor and filter of the augmented signal and the augmented noise: $\widehat{\mathbf{x}}_{k/s}$ and $\widehat{{\mathbf{v}}}^{\ast}_{k/s}$,  $s\leq k$.}

\vskip.25cm
Reasoning as above, it is proven that  $ {\mathcal S}^{\mathbf{x}}_{k,h}= \mathbb{A}_k\Psi^\mathbf{x}_h, \ h\leq k,$ and the following expression for the augmented signal estimators is obtained:
\begin{equation}\label{augmented_signal_pred_and_filter}
\widehat{\mathbf{x}}_{k/s}=\mathbb{A}_k e^\mathbf{x}_s, \ \ 1\leq s\leq k.
\end{equation}

Now, by using (\ref{observ_predictor}), the coefficients ${\mathcal{S}}^{\mathbf{v}^{\ast}}_{k,h}=E[\mathbf{v}^{\ast}_k\mu^{T}_h]$ can be expressed as
$$
{\mathcal{S}}^{\mathbf{v}^{\ast}}_{k,h}=
E\big[\mathbf{v}^{\ast}_k \mathbf{y}_{h}^T\big]-(1-\overline{\lambda}_h)\Big(E\big[
 \mathbf{v}^{\ast}_k\widehat{\mathbf{x}}^{T}_{h/h-1}\big]\overline{\mathcal{H}}_h^T + E\big[\mathbf{v}^{\ast}_k\widehat{{\mathbf{v}}}^{\ast T}_{h/h-1}\big]
 \Big), \ \ 1\leq h \leq k.
$$
From Proposition \ref{proposition 1}(b) and Proposition \ref{proposition 2}, it is clear that
$$E\big[\mathbf{v}^{\ast}_k\mathbf{y}^{T}_{h}\big] = (1-\overline{\lambda}_h)\Delta_k\Upsilon_h^{T}, \ \ 1\leq h \leq k,$$
and, in view of (\ref{general expression estimators mathbfx mathbfdotv}), we can write
$$E\big[\mathbf{v}^{\ast}_k\widehat{\mathbf{x}}^{T}_{h/h-1}\big] =\displaystyle{\sum_{j=1}^{h-1}}{\mathcal {S}}^{\mathbf{v}^{\ast}}_{k,j}\Pi^{-1}_j {\mathcal S}_{h,j}^{{\mathbf{x}}T}, \ \ \ \
  E\big[\mathbf{v}^{\ast}_k\widehat{{\mathbf{v}}}^{\ast T}_{h/h-1}\big] =\displaystyle{\sum_{j=1}^{h-1}}{\mathcal {S}}^{\mathbf{v}^{\ast}}_{k,j}\Pi^{-1}_j {\mathcal S}_{h,j}^{\mathbf{v}^{\ast }T }, \ \ \ \ h\geq 2.$$
As a consequence,
$$
{\mathcal S}^{\mathbf{v}^{\ast}}_{k,h}\hskip -0.05cm =
(1-\overline{\lambda}_h)\Big[ \ \Delta_k\Upsilon_h^T  -(1-\delta_{h,1})\displaystyle{\sum_{j=1}^{h-1}}{\mathcal S}^{\mathbf{v}^{\ast}}_{k,j}\Pi^{-1}_j \big( \overline{\mathcal{H}}_h{\mathcal S}^{\mathbf{x}}_{h,j} + {\mathcal S}_{h,j}^{\mathbf{v}^{\ast}} \big)^T \Big] ,\  \ h\geq 1,
$$
so, we can write
${\mathcal S}^{\mathbf{v}^{\ast}}_{k,h}= \Delta_k\Psi^{\mathbf{v}^{\ast}}_h, \ h\leq k$,
in which $\Psi^{\mathbf{v}^{\ast}}_h$ is a function satisfying
\begin{equation}\label{Psimathbf_dotv}
\Psi^{\mathbf{v}^{\ast}}_h=(1-\overline{\lambda}_h)\Big[\ \Upsilon_h^T  -(1-\delta_{h,1})\displaystyle{\sum_{j=1}^{h-1}}\Psi^{\mathbf{v}^{\ast}}_{j}\Pi^{-1}_j  \big( \overline{\mathcal{H}}_h{\mathcal S}^{\mathbf{x}}_{h,j} + {\mathcal S}_{h,j}^{\mathbf{v}^{\ast}} \big)^T \Big] ,\  \ h\geq 1.
\end{equation}
Hence, if we define
\begin{equation}\label{e_dotv}
e^{\mathbf{v}^{\ast}}_k = \displaystyle{\sum_{h=1}^{k}}\Psi^{\mathbf{v}^{\ast}}_h\Pi^{-1}_h\mu_h,\ \ k\geq 1; \ \ \ e^{\mathbf{v}^{\ast}}_0 = 0,
\end{equation}
and we use (\ref{general expression estimators mathbfx mathbfdotv}), it is concluded that
\begin{equation}\label{noise_pred_and_filter}
\widehat{{\mathbf{v}}}^{\ast}_{k/s}=\Delta_k e^{\mathbf{v}^{\ast}}_s, \ \ 1\leq s\leq k.
\end{equation}

\vskip.25cm

\noindent $\bullet$  \emph{Derivation of the filtering formulas: expressions (\ref{filter})-(\ref{innovation covariance matrix}).}

\vskip.25cm
In what follows, for the sake of simplicity, we will denote:
\begin{equation*}
  \mathbf{e}_k=\left(
\begin{array}{c}
 e^\mathbf{x}_k \\ e^{\mathbf{v}^{\ast}}_k
\end{array}
\right), \ \ k\geq 0;\ \ \  \mathbf{\Psi}_k=\left(
\begin{array}{c}
 \Psi^\mathbf{x}_k \\ \Psi^{\mathbf{v}^{\ast}}_k
\end{array}
\right), \ \ k\geq 1,
\end{equation*}
from which expression (\ref{filter}) for the filter is immediate, just using (\ref{signal_pred_and_filter}). Using the OPL, the filtering error covariances are expressed as $\widehat{\Sigma}_{k/k}= E[x_kx^T_k] - E[\widehat{x}_{k/k}\widehat{x}^T_{k/k}]$; so, using \emph{(H1)} and defining $\mathbf{T}_k=E[\mathbf{e}_k\mathbf{e}^{T}_k]$, expression (\ref{Filtering_error_cov}) is directly obtained.

The recursive relation (\ref{recursion_mathbfe_k}) for $\mathbf{e}_k$ is easily derived from (\ref{ex}) and (\ref{e_dotv}). Then, using that $\mathbf{e}_{k-1}$ is orthogonal to  $\mu_k$,
the recursion (\ref{recursion_mathbfT_k}) for $\mathbf{T}_k$ is straightforward.

In view of (\ref{observ_predictor}), (\ref{augmented_signal_pred_and_filter}) and (\ref{noise_pred_and_filter}), it is clear that
\begin{equation}\label{observ_predictor_demo}
\widehat{\mathbf{y}}_{k/k-1}=(1-\overline{\lambda}_k)\left( \overline{\mathcal{H}}_k\mathbb{A}_k \ | \  \Delta_k\right)\mathbf{e}_{k-1}, \ \ k\geq  1,
\end{equation}
which yields  (\ref{innovation}) for the innovation $\mu_k = \mathbf{y}_k- \widehat{\mathbf{y}}_{k/k-1}$.
To obtain its covariance matrix, $\Pi_k=E[\mu_k \mu^T_k]$, we just observe that the OPL  guarantees that $\Pi_k= \Sigma_k^{\mathbf{y}}- E[\widehat{\mathbf{y}}_{k/k-1}\widehat{\mathbf{y}}^T_{k/k-1}]$, and using (\ref{observ_predictor_demo}), expression (\ref{innovation covariance matrix}) for $\Pi_k$ is easily proven.

Expression (\ref{mathbfPsi})
 is derived just by combining (\ref{brevePsimathbfx}) and (\ref{Psimathbf_dotv}), using  in them that $ {\mathcal S}^{\mathbf{x}}_{h,j}= \mathbb{A}_h\Psi^\mathbf{x}_j$ and ${\mathcal S}^{\mathbf{v}^{\ast}}_{h,j}= \Delta_h\Psi^{\mathbf{v}^{\ast}}_j$ to obtain
$$
\mathbf{\Psi}_k=(1-\overline{\lambda}_k)\Big( \left( \overline{\mathcal{H}}_k\mathbb{B}_k \ | \  \Upsilon_k\right)-\left( \overline{\mathcal{H}}_k\mathbb{A}_k \ | \  \Delta_k \right)\displaystyle{\sum_{h=1}^{k-1}}\mathbf{\Psi}_h\Pi^{-1}_h \mathbf{\Psi}^T_h \Big)^T , \ \ k\geq 1,
$$
and taking into account that $\mathbf{T}_k= E[\mathbf{e}_k \mathbf{e}^T_k]=\displaystyle{\sum_{h=1}^{k}}\mathbf{\Psi}_h\Pi^{-1}_h \mathbf{\Psi}^T_h,\ k\geq 1.$

\vskip.25cm

\noindent $\bullet$  \emph{Derivation of the fixed-point smoothing formulas: (\ref{smoother})-(\ref{mathbfM_smoothing}).}

\vskip.25cm

The recursive relation (\ref{smoother}) for the fixed-point smoothers yields directly from (\ref{general expression estimators}) and, from it, the recursion for the error covariance matrices (\ref{Smoothing_error_cov}) is immediately obtained using the OPL.

In order to calculate the coefficients
$${\mathcal{\breve{S}}}^{x}_{k,k+N}=E[x_k\mu^{T}_{k+N}]= E\big[x_k\mathbf{y}_{k+N}^T\big]-E\big[x_k\widehat{\mathbf{y}}_{k+N/k+N-1}^T\big],$$
let us note that, from Proposition \ref{proposition 1}(b) and Proposition \ref{proposition 2},
$$
\begin{array}{ll}
  E\big[x_k\mathbf{y}^{T}_{k+N}\big] =  & (1-\overline{\lambda}_{k+N})\mathbb{\breve{B}}_k\mathbb{A}_{k+N}^{T}\overline{\mathcal{H}}_{k+N}^T
\end{array}
$$
Using (\ref{observ_predictor_demo}) and denoting $\mathbf{M}_{k,k+N}=E[x_k\mathbf{e}^{T}_{k+N}], \ N\geq 0$, it is clear that
$$E\big[x_k\widehat{\mathbf{y}}_{k+N/k+N-1}^T\big]=(1-\overline{\lambda}_{k+N})\mathbf{M}_{k,k+N-1} \left( \overline{\mathcal{H}}_{k+N}\mathbb{A}_{k+N} \ | \  \Delta_{k+N}\right)^T
$$
and, combining both expressions, (\ref{mathcal_breveS_smoothing}) is straightforward.

Finally, using the recursive relation (\ref{recursion_mathbfe_k}) for $\mathbf{e}_{k+N}$, expression (\ref{mathbfM_smoothing}) for the matrices  $\mathbf{M}_{k,k+N}$ is directly obtained; its initial condition is also easily derived by observing that, from the OPL, $\mathbf{M}_{k}=E[\widehat{x}_{k/k}\mathbf{e}^{T}_{k}]$ and using (\ref{filter}).

 \section{Linear estimation algorithms}
 \label{appendix linear estimation algorithm}

In this appendix, we present a recursive algorithm to obtain the
linear filter  and fixed-point smoother; its derivation, via an innovation approach, is analogous to
that of Theorem 2.

\vskip.25cm
{\em
 The LS linear filter, $\widehat{x}^L_{k/k}$, and the error covariance matrices, $\widehat{\Sigma}^L_{k/k}$, are obtained by
  $$
    \begin{array}{l}
        \widehat{x}^L_{k/k}=(A_k\ | \ 0)\mathbf{e}^L_k, \ \ k\geq 1,
    \\
        \widehat{\Sigma}^L_{k/k}=A_kB_k^T-(A_k \ | \ 0)\mathbf{T}^L_k(A_k \ | \ 0)^T, \ \ k\geq  1,
    \end{array}
    $$
  where the vectors $\mathbf{e}^L_k$ and the matrices $\mathbf{T}^L_k$  satisfy
   $$ \begin{array}{l}
      \mathbf{e}^L_k=\mathbf{e}^L_{k-1}+\mathbf{\Psi}^L_k\Xi^{-1}_k\eta_k,\ \ k\geq 1; \ \ \mathbf{e}^L_0=0,
   \\
        \mathbf{T}^L_k=\mathbf{T}^L_{k-1}+\mathbf{\Psi}^L_k\Xi^{-1}_k (\mathbf{\Psi}_k^{L})^{T}, \ \ k\geq 1; \ \ \mathbf{T}^L_0=0,
       \end{array}
    $$
    in which the matices $\mathbf{\Psi}^L_k$ are calculated from
    $$ \mathbf{\Psi}^L_k=(1-\overline{\lambda}_k)\Big( \left( \overline{H}_kB_k \ | \  \mathbf{F}_k\right)-\left( \overline{H}_kA_k \ | \  \mathbf{D}_k \right)\mathbf{T}^L_{k-1} \Big)^T , \ \ k\geq 1.$$
 The innovations $\eta_k$  and their covariance matrices, $\Xi_k$, are  calculated by
  $$  \begin{array}{l}
      \eta_k=y_k-(1-\overline{\lambda}_k)\left( \overline{H}_k A_k \ | \  \mathbf{D}_k\right)\mathbf{e}^L_{k-1}, \ \ k\geq  1.
   \\
   \Xi_k=   \Sigma_k^{y}-(1-\overline{\lambda}_k)^2\left( \overline{H}_kA_k \ | \  \mathbf{D}_k \right)\mathbf{T}^L_{k-1}   \left( \overline{H}_kA_k \ | \  \mathbf{D}_k\right)^T ,  \ \ k\geq  1,
    \end{array}
    $$
  where $\Sigma^{y}_k =(1-\overline{\lambda}_k)\left(E[H_{k}A_kB^{T}_kH^T_{k}]+\mathbf{D}_k \mathbf{F}_k^T\right)+\overline{\lambda}_k \Sigma^{w}_{k},\quad k\geq 1$.

  \vskip.25cm

The LS linear fixed-point smoothers, $\widehat{x}^L_{k/k+N}$, and
their error covariances, $\widehat{\Sigma}^L_{k/k+N}$, are recursively obtained by
  $$  \begin{array}{l}
        \widehat{x}^L_{k/k+N}=\widehat{x}^L_{k/k+N-1}+ \mathcal{S}^{L}_{k,k+N}\Xi^{-1}_{k+N}\eta_{k+N}\!\!\!,  \ N\geq 1,
    \\
        \widehat{\Sigma}^L_{k/k+N}=\widehat{\Sigma}^L_{k/k+N-1}-\mathcal{S}^{L}_{k,k+N}\Xi^{-1}_{k+N}(\mathcal{S}^{L}_{k,k+N})^T, \ \ N\geq 1,
    \end{array}$$
where
  $$
      {\mathcal{S}}^{L}_{k,k+N}=(1-\overline{\lambda}_{k+N})\left(\big(B_k \ | \  0\big)-\mathbf{M}^L_{k,k+N-1}\right)
 \left( \overline{H}_{k+N}A_{k+N} \ | \  \mathbf{D}_{k+N}\right)^T, \ \ N\geq 1,$$
with $$  \begin{array}{l}
          \mathbf{M}^L_{k,k+N}=\mathbf{M}^L_{k,k+N-1}+\mathcal{S}^{L}_{k,k+N}\Xi^{-1}_{k+N}(\mathbf{\Psi}^{L}_{k+N})^T, \ \ N\geq 1;\\  \mathbf{M}^L_{k}=  (A_k \ | \ 0)\mathbf{T}^L_k, \ \ k\geq 1.
        \end{array}$$

     }


\begin{thebibliography}{00}


\bibitem{Singh_et_al_Springer_2022}
U. Singh, A. Abraham, A. Kaklauskas, T. Hong, \emph{Smart Sensor Networks. Analytics, Sharing and Control},
Springer, Switzerland, 2022.
https://doi.org/10.1007/978-3-030-77214-7.

\bibitem{Liu_et_al_Springer_2019}
Q. Liu, Z. Wang, X. He, \emph{Stochastic Control and Filtering over Constrained Communication Networks},
Springer, Switzerland, 2019.
https://doi.org/10.1007/978-3-030-00157-5.


\bibitem{Yang_2016}
     Y. Yang, Y. Liang, Q. Pan, Y. Qin,  F. Yang,
     Distributed fusion estimation with square-root array implementation for Markovian jump linear systems with random parameter matrices and cross-correlated noises,
     {\em Inf. Sci.} 370--371 (2016) 446--462.
     https://doi.org/10.1016/j.ins.2016.08.020.

\bibitem{Wang_2017}
    W. Wang, J. Zhou,
    Optimal linear filtering design for discrete time systems with cross-correlated stochastic parameter matrices and noises,
    {\em IET Control Theory Appl.} 11 (2017), 3353--3362.
    https://doi.org/10.1049/iet-cta.2017.0425.


\bibitem{Han_2018}
     F. Han, H. Dong, Z. Wang, G. Li, F.E. Alsaadi,
     Improved Tobit Kalman filtering for systems with random parameters via conditional expectation,
     {\em  Signal Process.} 147 (2018)  35--45.
     http://dx.doi.org/10.1016/j.sigpro.2018.01.015.


\bibitem{Caballero_DSP_2019}
    R. Caballero-\'{A}guila, A. Hermoso-Carazo, J. Linares-P\'{e}rez
    Centralized filtering and smoothing algorithms from outputs with random parameter matrices transmitted through uncertain communication channels,
    {\em Digit. Signal Process.} 85 (2019) 77--85.
    https://doi.org/10.1016/j.dsp.2018.11.010


\bibitem{Liu_2020}
    W. Liu, X. Xie, W. Qian, X. Xu, Y. Shi,
    Optimal linear filtering for networked control systems with random matrices, correlated noises, and packet dropouts,
    {\em IEEE Access}. 8 (2020) 59987--59997.
    http://dx.doi.org/10.1109/ACCESS.2020.2983122.

\bibitem{Sun_IEEE_TSP_2020}
    S. Sun,
    Distributed optimal linear fusion predictors and filters for systems with random parameter matrices and correlated noises,
    {\em IEEE Trans. Signal Process.} 68 (2020) 1064--1074.
    https://doi.org/10.1109/TSP.2020.2967180.

\bibitem{Caballero_IJSS_2023}
    R. Caballero-Águila, J. Linares-Pérez,
    Distributed fusion filtering for uncertain systems with coupled noises, random delays and packet loss prediction compensation,
    {\em Int. J. Syst. Sci.} 54 (2023) 371--390.
    https://doi.org/10.1080/00207721.2022.2122905.





\bibitem{Li_2017}
     W. Li, Y. Jia, J. Du,
     Distributed filtering for discrete-time linear systems with fading measurements and time-correlated noise,
     {\em Digit. Signal Process.} 60 (2017) 211--219.
     https://doi.org/10.1016/j.dsp.2016.10.003.


\bibitem{Liu_2017}
     W. Liu, P. Shi, J.S. Pan,
     State estimation for discrete-time Markov jump linear systems with time-correlated and mode-dependent measurement noise,
     {\em Automatica}. 85 (2017) 9--21.
     https://doi.org/10.1016/j.automatica.2017.07.025.


\bibitem{Geng_2019}
     H. Geng, Z. Wang, Y. Cheng, F. Alsaadi, A.M. Dobaie,
     State estimation under non-Gaussian L{\'e}vy and time-correlated additive sensor noises: A modified Tobit Kalman filtering approach,
     {\em  Signal Process.} 154 (2019)  120--128.
     https://doi.org/10.1016/j.sigpro.2018.08.005.


\bibitem{Liu_2019}
    W. Liu, P. Shi,
    Convergence of optimal linear estimator with multiplicative and time-correlated additive measurement noises,
    {\em IEEE Trans. Autom. Control}  64 (2019) 2190--2197.
    https://doi.org/10.1109/TAC.2018.2869467.


\bibitem{Caballero_INFFUS_2020}
    R. Caballero-\'Aguila, A. Hermoso-Carazo, J. Linares-P\'erez,
    Networked fusion estimation with multiple uncertainties and time-correlated channel noise,
    {\em Inf. Fusion} 54 (2020) 161--171.
    https://doi.org/10.1016/j.inffus.2019.07.008.



\bibitem{Ma_SIGPRO_2020}
    J. Ma, S. Sun, 2020.
    Optimal linear recursive estimators for stochastic uncertain systems with time-correlated additive noises and packet dropout compensations,
    {\em Signal Process.} 176, 107704.
    https://doi.org/10.1016/j.sigpro.2020.107704.


\bibitem{Caballero_SENSORS_2022}
    R. Caballero-Águila, J. Hu, J. Linares-Pérez, 2022.
    Two Compensation Strategies for Optimal Estimation in Sensor Networks with Random Matrices, Time-Correlated Noises, Deception Attacks and Packet Losses,
    \emph{Sensors} 22, 8505.
    https://doi.org/10.3390/s22218505.





\bibitem{Mahmoud_2019}
M.S. Mahmoud, M.M. Hamdan, U.A. Baroudi,
Modeling and control of Cyber-Physical Systems subject to cyber attacks: A survey of recent advances and challenges,
{\em Neurocomputing}. 338 (2019) 101--115.
https://doi.org/10.1016/j.neucom.2019.01.099.

\bibitem{Sanchez_2019}
H.S. Sánchez, D. Rotondo, T. Escobet, V. Puig, J. Quevedo,
Bibliographical review on cyber attacks from a control oriented perspective,
{\em Annu. Rev. Control}. 48 (2019) 103--128.
https://doi.org/10.1016/j.arcontrol.2019.08.002.


\bibitem{Wang_2018_IJFI}
Z. Wang, D. Wang, B. Shen, F.E. Alsaadi,
Centralized security-guaranteed filtering in multirate-sensor fusion under deception attacks,
{\em J. Frankl. Inst.} 355 (2018) 406–420.
https://doi.org/10.1016/j.jfranklin.2017.11.010.

\bibitem{Han_2019}
F. Han, H. Dong, Z. Wang, G. Li,
Local design of distributed H$_\infty$-consensus filtering over sensor networks under multiplicative noises and deception attacks,
{\em Int. J. Robust Nonlinear Control.}  29 (2019) 2296--2314.
https://doi.org/10.1002/rnc.4493.

{
\bibitem{Qu_2022}
F. Qu, E. Tian, X. Zhao, 2022.
Chance-Constrained $H_\infty$ State Estimation for Recursive Neural Networks Under Deception Attacks and Energy Constraints: The Finite-Horizon Case,
{\em IEEE Trans. Neural Netw. Learn. Syst}. 
https://doi.org/10.1109/TNNLS.2021.3137426.
}

\bibitem{Yang-et-al-2019-Automatica}
W. Yang, Y. Zhang, G. Chen, C. Yang, L. Shi,
Distributed filtering under false data injection attacks,
{\em Automatica.}  102 (2019)  34--44.
https://doi.org/10.1016/j.automatica.2018.12.027.


\bibitem{Caballero_SENSORS_2020}
R. Caballero-\'Aguila, A. Hermoso-Carazo, J. Linares-P\'erez, 2020.
A two-phase distributed filtering algorithm for networked uncertain systems with fading measurements under deception attacks,
{\em Sensors}. 20, 6445.
https://doi.org/10.3390/s20226445.


\bibitem{Xiao-et-al-2020}
S. Xiao, Q. Han, X. Ge, Y. Zhang,
Secure distributed finite-time filtering for positive systems over sensor networks under deception attacks,
{\em IEEE Trans. Cybern.} 50 (2020) 1200--1228.
https://doi.org/10.1109/tcyb.2019.2900478.




\bibitem{Ma-Sun-Sensors_2023}
Y. Ma, S. Sun, 2023.
Distributed Optimal and Self-Tuning Filters Based on Compressed Data for Networked Stochastic Uncertain Systems with Deception Attacks,
{\em Sensors.} 23, 335.
https://doi.org/10.3390/s23010335.

\bibitem{Ma-et-al-IEEE_TSIPN_2021}
L. Ma, Z. Wang, Y. Chen, X. Yi,
Probability-guaranteed distributed secure estimation for nonlinear systems over sensor networks under deception attacks on innovations,
{\em IEEE Trans. Signal Inf. Proc. Netw.} 7 (2021) 465--477.
https://doi.org/10.1109/TSIPN.2021.3097217.




\bibitem{Zhao_Zhang-Neurocomputing2016}
    H. Zhao, C. Zhang,
    Non-Gaussian noise quadratic estimation for linear discrete-time time-varying systems,
    \emph{Neurocomputing.} 174 (2016) 921--927.
    http://dx.doi.org/10.1016/j.neucom.2015.10.015.


\bibitem{Caballero_et_al_AMC2016}
    R. Caballero-Águila, I. García-Garrido, J. Linares-Pérez,
    Quadratic estimation problem in discrete-time stochastic systems with random parameter matrices,
    \emph{Appl. Math. Comp.} 273 (2016) 308-–320.
    http://dx.doi.org/10.1016/j.amc.2015.10.005


\bibitem{Cacace_et_al-Automatica2017}
    F. Cacace, F. Conte, A. Germani, G. Palombo,
    Feedback quadratic filtering,
    \emph{Automatica.} 82 (2017) 158--164.
    https://doi.org/10.1016/j.automatica.2017.04.046.


\bibitem{Li_et_al-Complexity2020}
    L. Li, L. Tan, X. Song, X. Yan, 2020.
    Quadratic Filtering for Discrete-Time Systems with Measurement Delay and Packet Dropping,
    \emph{Complexity.} 2020, 1725121.
    https://doi.org/10.1155/2020/1725121.

\bibitem{Liu_et_al-Automatica2020}
   Q. Liu, Z. Wang, Q. Han, C. Jiang, 2020.
   Quadratic estimation for discrete time-varying non-Gaussian systems with multiplicative noises and quantization effects,
  \emph{ Automatica.} 113, 108714.
  https://doi.org/10.1016/j.automatica.2019.108714.

\bibitem{Wang_et_al-IEEETSP2022}
  S. Wang, Z. Wang, H. Dong, Y. Chen, F.E. Alsaadi,
  Recursive Quadratic Filtering for Linear Discrete Non-Gaussian Systems Over Time-Correlated Fading Channels,
  \emph{IEEE Trans. Signal Process.} 70 (2022) 3343--3356.
  https://doi.org/10.1109/TSP.2022.3182511.

\bibitem{Wang_et_al-Automatica2023}
  S. Wang, Z. Wang, H. Dong, Y. Chen, 2023.
  Recursive state estimation for stochastic nonlinear non-Gaussian systems using energy-harvesting sensors: A quadratic estimation approach,
  \emph{Automatica.} 147, 110671.
  https://doi.org/10.1016/j.automatica.2022.110671.

{
\bibitem{Shmaliy2020}
 Y.S. Shmaliy, S. Zhao, C.K. Ahn,
 Kalman and UFIR state estimation with coloured measurement noise using backward Euler method,
 \emph{IET Signal Process}. 14 (2020) 64--71.
https://doi.org/10.1049/iet-spr.2019.0166.}

{
\bibitem{Wang2023}
X. Wang, E. Tian, B. Wei, J. Liu,
Novel attack-defense framework for nonlinear complex networks: An important-data-based method,
\emph{Int J Robust Nonlinear Control}, 33 (2023) 2861–2878.
https://doi.org/10.1002/rnc.6551
}


{
\bibitem{Liu2023}
Q. Liu, S. Zheng, Z. Wang, N. D. Alotaibi, F. E. Alsaadi,
Distributed estimation for multi-agent systems with relative measurements and quantized communication: A feedback quadratic framework,
\emph{Int J Robust Nonlinear Control}, 33 (2023) 3164–3184.
 https://doi.org/10.1002/rnc.6564
}

\end{thebibliography}
\end{document}